\RequirePackage{etoolbox}
\pretocmd\PackageWarning{%
  \edef\pkgname{#1}\edef\hyperrefname{hyperref}%
  \ifx\pkgname\hyperrefname
    \expandafter\gobblethree
  \fi
}{}{\undefined}
\newcommand*{\gobblethree}[3]{}

\documentclass[preprint,journal]{vgtc}       % preprint (journal style)

\ifpdf%                                % if we use pdflatex
  \pdfoutput=1\relax                   % create PDFs from pdfLaTeX
  \usepackage{graphicx}                % allow us to embed graphics files
  \DeclareGraphicsExtensions{.pdf,.png,.jpg,.jpeg} % for pdflatex we expect .pdf, .png, or .jpg files
\else%                                 % else we use pure latex
  \usepackage{graphicx}                % allow us to embed graphics files
  \DeclareGraphicsExtensions{.eps}     % for pure latex we expect eps files
\fi%

\graphicspath{{figures/}{pictures/}{images/}{./}} % where to search for the images

\usepackage{microtype}                 % use micro-typography (slightly more compact, better to read)
\PassOptionsToPackage{warn}{textcomp}  % to address font issues with \textrightarrow
\usepackage{textcomp}                  % use better special symbols
\usepackage{mathptmx}                  % use matching math font
\usepackage{times}                     % we use Times as the main font
\usepackage{cite}                      % needed to automatically sort the references
\usepackage{tabu}                      % only used for the table example
\usepackage{booktabs}                  % only used for the table example
\usepackage{enumitem}
\usepackage{lipsum}

\usepackage[usenames,dvipsnames,svgnames,table]{xcolor}

\onlineid{1124}

\vgtccategory{Research}
\vgtcpapertype{Empirical Study}

\title{An Examination of Grouping and Spatial Organization Tasks for\\High-Dimensional Data Exploration}

\author{John Wenskovitch, \textit{Member, IEEE}, and Chris North}
\authorfooter{
\item
 John Wenskovitch and Chris North are with the Virginia Tech Department of Computer Science and the Discovery Analytics Center. E-mails: $\{$jw87 $|$ north$\}$@cs.vt.edu.  John Wenskovitch is also currently employed by Pacific Northwest National Laboratory.
}

\abstract{How do analysts think about grouping and spatial operations?  This overarching question incorporates a number of points for investigation, including understanding how analysts begin to explore a dataset, the types of grouping/spatial structures created and the operations performed on them, the relationship between grouping and spatial structures, the decisions analysts make when exploring individual observations, and the role of external information.  This work contributes the design and results of such a study, in which a group of participants are asked to organize the data contained within an unfamiliar quantitative dataset.  We identify several overarching approaches taken by participants to design their organizational space, discuss the interactions performed by the participants, and propose design recommendations to improve the usability of future high-dimensional data exploration tools that make use of grouping (clustering) and spatial (dimension reduction) operations.%
} % end of abstract


\CCScatlist{ % not used in journal version
 \CCScat{K.6.1}{Management of Computing and Information Systems}%
{Project and People Management}{Life Cycle};
 \CCScat{K.7.m}{The Computing Profession}{Miscellaneous}{Ethics}
}

\vgtcinsertpkg

\begin{document}


\firstsection{Introduction}

\maketitle

Sensemaking refers to a cognitive process for acquiring, representing, and organizing information in order to address a task, solve a problem, or make a decision~\cite{klein2006making,russell1993cost}.  A number of models with varying levels of information granularity have been proposed for approaching and solving sensemaking problems~\cite{pirolli1999information,pirolli2005sensemakingprocess,russell1993cost}.  These models represent strategies for addressing a variety of sensemaking problems.  For example, Pirolli and Card's Sensemaking Process~\cite{pirolli2005sensemakingprocess} %(see Figure~\ref{fig:sensemaking}) 
is designed for sensemaking problems faced by intelligence analysts.  Despite the specific challenge addressed by each of these models, they all highlight the need to organize the data.  Continuing the intelligence analyst example, they may work to understand the actors and motivations by grouping documents by location, by person, or by subplot.  

% \begin{figure}[!tb]
%   \centering
%   \includegraphics[width=0.9\columnwidth]{figures/sensemakingloop.JPG}
%   \caption{In the Sensemaking Process~\cite{pirolli2005sensemakingprocess}, intelligence analysts transform raw information into reportable results through organizational stages that filter, extract, and structure data.}
%   \label{fig:sensemaking}
% \end{figure}

A fundamental behavior in sensemaking is the act of grouping similar observations in order to understand their properties, effectively forming a cluster.  This organizational strategy is true both in paper-based sensemaking tasks~\cite{drucker2011helping,whittaker2001character} and in tasks performed on electronic displays~\cite{andrews2010space,endert2012semantics}.  Clusters therefore have a natural connection to sensemaking.  Clusters can also help to reduce clutter in a workspace, compressing similar observations into a group that requires less physical or screen space~\cite{mander1992pile,rose1993content}.  Simplifying the workspace leads to further cognitive benefits, as humans struggle to think about more than a small number of observations or dimensions at one time~\cite{self2018observation}. Thus, using groups of items to perform analysis tasks can lead to improved memory and recall by providing a simplified method of understanding the data~\cite{curiel1998mental}.  Previous research has shown that humans use a variety of organizational principles to cluster information~\cite{dourish1999building}, even when addressing the same task~\cite{andrews2010space}.  In order to identify clusters computationally, hundreds of clustering algorithms have been implemented, each with strengths and weaknesses~\cite{estivillcastro2002somany}.  

Another technique for sensemaking, particularly relevant to multidimensional datasets, is to embed the high-dimensional data in a low-dimensional projection in such a way that the structure of the high-dimensional space (e.g., clusters and outliers) is maintained in the projection~\cite{ingram2009glimmer,maaten2008visualizing,torgerson1958theory}.  Often, these projections embody a ``proximity$\approx$similarity'' metaphor, in which the distance between observations represents the similarity of those observations.  As a result, groups of similar observations form clusters within the projection.  A number of interactive tools make use of such techniques for exploratory data analysis and sensemaking~\cite{bradel2014multimodel,dowling2018sirius,endert2012semantic,paulovich2011piece,self2018observation}. 

The intersection of these grouping and spatialization sensemaking techniques (and their corresponding clustering and dimension reduction algorithms) has become an area of interest for visualization research, including tools and techniques~\cite{ding2007adaptive,lee2012ivisclustering,ng2001spectral,zha2002spectral}, studies~\cite{brehmer2014interviews,saket2014node}, and surveys~\cite{endert2017state,wenskovitch2018towards}.  However, the behavior of users when performing these tasks on a dataset has not been thoroughly explored, with the most similar study to this work performing a post-hoc analysis of created groups after the sensemaking task~\cite{endert2012semantics}.  In contrast, this work presents a study in which participants have been asked to perform spatial and grouping operations within a single sensemaking task, with the goal of understanding the organizational strategies of the participants and the interplay between the operations.

In particular, we note the following contributions:
\begin{enumerate}[noitemsep,nolistsep]
    \item The design and execution of a study to understand the cognitive group and spatialization processes involved in organizing a dataset for a sensemaking task.
    \item A discussion of organizational strategies and structures, grouping and spatial operations, decision making, and external knowledge displayed by study participants.
    \item Design recommendations that are intended for both current and future tools, with the goal of better supporting user organizational processes for sensemaking in interactive visualization systems.
\end{enumerate}

We identified three overarching organizational strategies demonstrated by participants, saw tight interplay between spatial and grouping operations during sensemaking, and noted the creation of organizational spaces that were more complex than those currently supported by interactive sensemaking tools.  The results from this study can be used to inform the design of future tools for interactive visual data exploration.

\section{Background and Related Work}

This work was primarily inspired by ``The Semantics of Clustering'' by Endert et al~\cite{endert2012semantics}.  In that study, participants were directed to perform a spatial sensemaking task in a large display space, organizing a collection of documents with only manual layout capabilities.  After completing the task and participating in a semi-structured interview, the participants were asked to draw the final cluster structure of their organizational space.  While this study provides a useful starting point for understanding the spatial and grouping behaviors of analysts, the post-hoc analysis of clusters misses information about the development of clusters and the interplay between grouping and spatial actions.  Indeed, research has demonstrated that categories that emerge during the analysis process are often shifting and ad-hoc, evolving throughout the course of the analysis to represent the information uncovered~\cite{barsalou1983ad,keil1992concepts}.  Our goal with the study discussed in this work is to investigate grouping and spatial behaviors throughout the analysis process.

The card-based approach used in this study was inspired by collaboration studies performed by Robinson~\cite{robinson2008collaborative} and Isenberg et al~\cite{isenberg2008exploratory}.  As noted by Robinson, the intention of this strategy was to permit understanding without constraints that may be imposed by software tools.  For example, there are a variety of methods by which participants indicated groups during our study, including by stacking observations, by overlapping observations, by positioning observations so that cards were touching, and by positioning observations relatively close to each other.  Similar studies also include affinity diagramming techniques for organization~\cite{pandey2016towards}.  Other possible metaphors for indicating groups could be colored labels or drawing boundaries around collections of observations.  Further, Sellen and Harper suggest that using physical artifacts in such studies can be useful for revealing how participants make use of their affordances to complete tasks, thereby providing interface and interaction design guidelines~\cite{sellen1997paper}.

\subsection{Sensemaking and Cognition}

Sensemaking is an iterative process, building up an internal representation of information in order to achieve the user's goal~\cite{russell1993cost}.  Distributed and embodied cognition share complementary roles in human sensemaking.  Distributed cognition refers to the idea that external spaces can be used to extend and support cognitive reasoning.  Analysts can thus use objects or symbols as a means of externally encoding relationships~\cite{rogers1994distributed}.  ``Space to Think'' demonstrated that space plays a meaningful role in sensemaking, providing a large high-resolution display grid to permit analysts to organize hypotheses and evidence spatially~\cite{andrews2010space}.  Spatial memory has also been leveraged by Robertson et al. for document arrangement in their Data Mountain system~\cite{robertson1998data}, and the role of spatial memory has been extended into 3D interfaces for information retrieval tasks by Cockburn and McKenzie~\cite{cockburn2002evaluating}.

Embodied cognition~\cite{wilson2002six} focuses on the integration of the physical body and the environment with internal resources, reflecting how the body influences cognition.  Embodied cognition allows analysts to offload cognition and create understanding within their workspace, allowing physical navigation to provide more meaning to locations~\cite{andrews2012analysts}.  ``Space to Think'' has been demonstrated to extend to more complex spaces which contain multiple displays and devices~\cite{chung2018savil,chung2014visporter,hamilton2014conductor}.

\subsection{Dimension Reduction and Clustering Tools} \label{sec:dr_clust}

Interactive dimension reduction (DR) and clustering are both active topics in visual analytics research.  In many of these systems designed to support user sensemaking and exploratory data analysis, a learning module responds to incremental user feedback, structuring and displaying subsequent visual representations of the data in a way that reflects the goals of the analyst exploring the data~\cite{shipman1999formality}.  One goal of this study is to better design future tools in this space by carefully examining user interactions and their motivations when applied to a dataset.

Interactions in high-dimensional data exploration systems can be organized into two broad categories: \textit{explanatory} and \textit{expressive}~\cite{endert2011observation}.  Explanatory interactions, or surface-level interactions, seek to understand the data without altering any underlying model.  Such interactions are often used to support low-level tasks such as finding extrema or retrieving a value~\cite{amar2005lowlevel}.  In contrast, expressive interactions communicate some intent from the analyst to the system, resulting in a model update.  Parametric interactions are directly applied to model parameters, such as changing the weight on a dimension in Andromeda~\cite{self2018observation}.  Observational-level interactions are applied to individual data items in a projection, which are then used to infer the analyst intent.

The semantic interaction work by Endert et al.~\cite{endert2012semantic} catalyzed much of the current research into expressive interactions in DR systems.  Tools that resulted from this research direction can be divided into those that support quantitative and text data, though text data must be converted into a numeric representation for the algorithms to process.  Quantitative tools such as Dis-Function~\cite{brown2012disfunction} present similarity-based projections of the data, with a user interest weight vector applied to the dimensions of the dataset to influence the projection in response to user interactions.  %Similar tools and learning approaches are seen in work by Joia et al.~\cite{joia2011lamp}, Paulovich et al.~\cite{paulovich2011piecewise}, Mamani et al.~\cite{mamani2013userdriven}, and Molchanov et al~\cite{molchanov2014interactive}.  
These learning techniques also support text data, as seen in systems like StarSPIRE~\cite{bradel2014multimodel,wenskovitch2018effect} and Cosmos~\cite{dowling2019interactive}.  One key change for the text case is creating a similarity computation based on a distance metric such as Cosine distance to resolve issues with term sparcity~\cite{singhal2001modern} or Gower distance to adjust for missing terms~\cite{gower1971general}.  In contrast, quantitative tools often make use of Euclidean distance~\cite{wenskovitch2019pollux,dowling2018sirius,self2018observation}, though other distance metrics are also seen in the literature.  With this study, we seek to both identify and understand which distances are important to the organizational structures created by the participants.

Interactive clustering supports similar data exploration processes, with analysts often seeking to find the clustering assignment that best suits their understanding or interpretation of the data~\cite{andrienko2009interactive,macinnes2010visual}.  The Semantic Interaction paradigm can also be applied to interactive clustering, with a goal of understanding these analysts and adapting the clustering to suit their intent~\cite{chuang2014human,guo2010interactive,sourina2005visual}.  Using the principle of ``Assignment Feedback'' as coined by Dubey et al.~\cite{dubey2010clusterlevel}, analysts are often afforded the ability to directly move observations between clusters~\cite{basu2010assisting,coden2017method}, thereby supplying constraints on future iterations of the cluster assignments.  Beyond direct interaction with observations, systems can also support direct interaction with the clusters themselves, including operations such as merging and splitting clusters~\cite{boudjeloudassala2016interactive,choo2013utopian,hoque2016interactive}, removing clusters~\cite{basu2010assisting,dobrynin2005sophia,guo2010interactive,macinnes2010visual}, and expanding clusters~\cite{andrienko2009interactive,basu2010assisting,boudjeloudassala2016interactive,dobrynin2005sophia}.  In this study, we identify the frequency of these cluster-oriented techniques and tie them to the exploratory intent of the user.

\subsection{Cognitive Dimension Reduction and Clustering}

Existing research in both the cognitive science and visualization communities has also explored the understanding of human factors in the realm of DR and clustering, with a particular focus on both perception and cognition of these techniques.  With respect to the cognitive science field, researchers are actively studying the effects of both spatializations and groupings.  Baylis and Driver~\cite{baylis1992visual} consider the role of proximity, finding that visual attention is not based solely on positional information.  Similarly, Kramer and Johnson~\cite{kramer1991perceptual} consider the influence of Gestalt grouping principles of similarity, closure, and proximity in attention-based tasks, noting that the inclusion of distractors within groups yields a negative performance effect.  Gillam~\cite{gillam2001varieties} discusses the role of grouping on spatial layout in great detail, proposing new Gestalt grouping principles such as common region and connectedness.  Others have examined specific areas in cognitive grouping, such as the work of Zadeh on fuzzy sets~\cite{zadeh1965fuzzy} and prototype theory~\cite{zadeh1996note}, Fisher's investigation of conceptual clustering~\cite{fisher1987knowledge,fisher1987improving}, and Duncan's set of superimposition~\cite{duncan1984selective}.

In the visualization community, Nonato and Aupetit examined the impact of distortions on analytical tasks performed by users when exploring projections of high-dimensional data~\cite{nonato2019multidimensional}.  Such distortions are common features of these projections, as the reduction from a high-dimensional space to a 2D (or occasionally 3D) space necessitates a loss of information.  Similarly, Lewis et al. explore the reliability of human embedding evaluations, seeking to understand whether analysts agree on the quality of a projection as well as what types of embedding structures are favored by analysts~\cite{lewis2012behavioral}.  In the realm of clustering, Lewis et al. also contribute a study that compares standard quality measures for clustering (e.g., Dunn's Index~\cite{dunn1974well} %, Silhouette Coefficient~\cite{rousseeuw1987silhouettes}, 
and Cali{\'n}ski-Harabasz~\cite{calinski1974dendrite}, among others) to the interpretations generated by analysts who evaluate the clustering assignments~\cite{lewis2012human}.  Sedlmair et al. introduce a taxonomy of visual cluster separation factors, examining the density, isotropy, and clumpiness of clusters in scatterplots and finding that visual inspection with various DR and clustering techniques failed to match data classifications in more than half of the human inspection cases~\cite{sedlmair2012taxonomy}.

\section{Experimental Design}

The overarching question that motivates this research focuses on the organizational processes of analysts when performing exploratory data analysis.  For the purpose of this study, we define an ``analyst'' as a person with expertise in managing complex data, more particularly high-dimensional quantitative data for the purpose of this study, as well as familiarity with basic data science techniques such as projections and clustering.  We wish to understand the cognitive processes that underlie the approach that analysts take when trying to find insight within an unfamiliar dataset.  For example, when analysts are only afforded grouping and spatialization actions, how will they organize a collection of observations?  Further, we wish to understand how analysts begin to explore a dataset, the types of group structures created and types of grouping operations performed, and the decisions that analysts make when exploring individual observations.  This study was designed to investigate these components of the exploratory data analysis process.

\subsection{Participants}

% \reviewer{R1:  As the authors note, it would be ideal to have more participants. That would simply give the reader more confidence in the findings and implications made. It's still a pretty small number of participants to be making generalizations from.}

% \reviewer{R2:  Although the framing of the paper suggests this is a study of analysts, in reality the participants are from engineering and technology fields (not enough detail is provided about participant demographics). While this does not invalidate the study, it does invalidate the claim that this study increases our understanding of how ``analysts think about grouping and spatial operations''.}

% \reviewer{R2:  I struggle to understand why the study participants were split into groups with labeled and unlabeled data. Although it did allow for some observations regarding the use of external knowledge, it seems to have added a lot of complexity to an already-complex study without directly stemming from the overarching research question. As this is not a controlled lab study, it is not strictly necessary to have more than one condition.}

We recruited 16 participants from statistics and computer science disciplines in an academic setting.  These participants all self-reported some degree of experience with either data science or exploratory data analysis, with this experience ranging from taking a course that included a data science component to developing new methods and tools for data analysis.  The participant cohort included three undergraduate students, eleven graduate students, and two faculty members.

\begin{figure}[!tb]
    \centering
    \includegraphics[width=\linewidth]{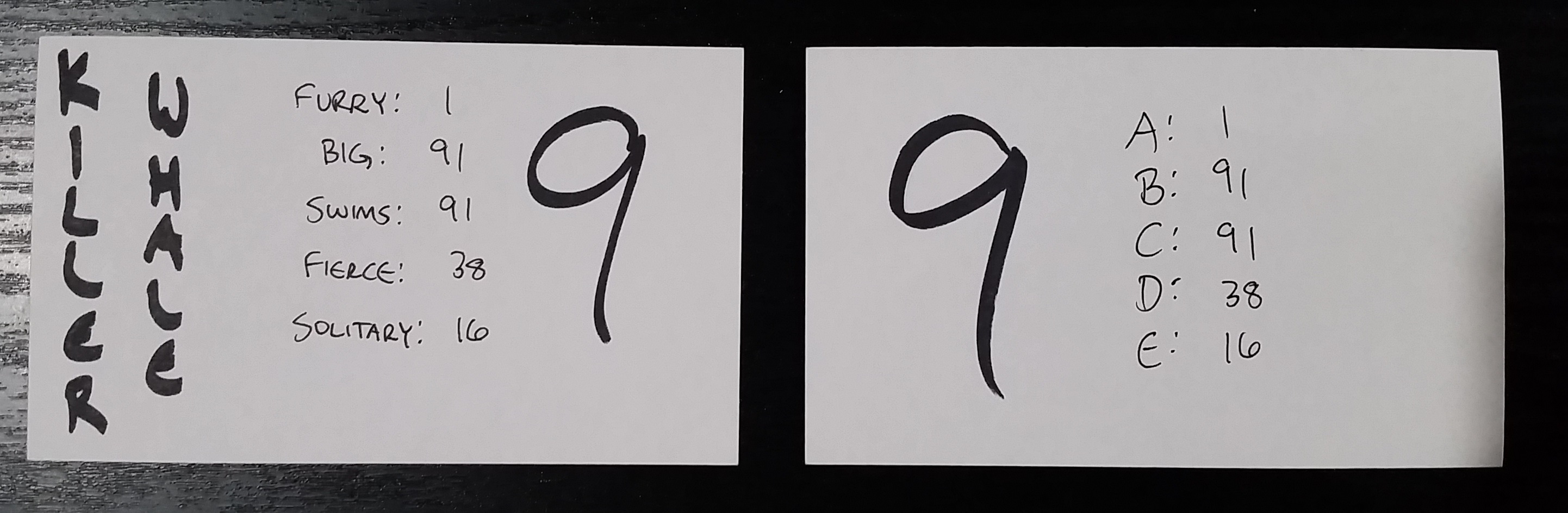}
    \caption{A photo of the Killer Whale card in both the \textit{(left)} labeled and \textit{(right)} abstract datasets.\vspace{-0.2cm}}
    \label{fig:cards}
\end{figure}

Using a between-subjects design, we divided the participants into two equal groups, each of which performed their organizational tasks on a separate set of observations.  The group that received the labeled dataset (participants referred to as L1--L8) received 17~index cards with an animal name and five dimensions that describe the animal (see Fig.~\ref{fig:cards} left).  The second group received the same data but in abstract form (see Fig.~\ref{fig:cards} right; participants referred to as A1--A8), with all animal-related contextual information removed from the cards.  In the original dataset, the attributes for the animals are normalized to a 0--100 range (the subset of animals and dimensions selected in the study dataset made the largest value on the cards~91).  The study dataset is provided in the supplemental material.  Participants were asked to ignore the large numbers on the cards; they were added for better video capture.

\subsection{Dataset}

The animals dataset provided to the participants is a reduced version of that created by Lampert et al~\cite{lampert2009animals}, selected because of its general knowledge applicability to all potential participants.  From the initial dataset, we rounded all decimal values to the nearest integer, and then reduced the number of animals and the number of dimensions.  We selected 17~animals with the foreknowledge that they could be naturally divided into three groups of five animals plus two outliers, but that alternate classifications and group assignments were possible.  Five dimensions were selected to make the task challenging but not overwhelmingly difficult, with a dimension selected to describe each of the three groups plus two additional noise dimensions.  One potential division of this dataset %(displayed in Figure~\ref{fig:my_organization} with additional hierarchical clustering) 
could be:
\begin{itemize}[noitemsep, nolistsep]
    \item \textbf{Predators} (described by Fierce): Bobcat, Grizzly Bear, Leopard, Polar Bear, Wolf
    \item \textbf{Aquatic} (described by Swims):  Blue Whale, Killer Whale, Otter, Seal, Walrus
    \item \textbf{Large Herbivores} (described by Big):  Cow, Deer, Giraffe, Hippopotamus, Moose
    \item \textbf{Outliers:}  Bat, Squirrel
\end{itemize}
%
% \begin{figure}[!tb]
%     \centering
%     \includegraphics[width=\linewidth]{figures/my_organization.png}
%     \caption{One of many potential organizational structures possible to generate from the study dataset. \todo{drop?}}
%     \label{fig:my_organization}
% \end{figure}
%
However, alternative natural groupings are possible.  For example, the Polar Bear and Hippopotamus have Swims attributes that could place them in the Aquatic group, or the Killer Whale could be a Predator (and the Bat has a similar Fierce attribute).  The animals could also be divided into two groups rather than three:  Aquatic and Non-Aquatic, or Furry and Non-Furry.

\subsection{Study Procedure}

Noting the cognitive strain experienced by participants during a pilot study, we elected to limit the length of the study to one hour in order to minimize fatigue and frustration effects.  Participants began the study by responding to four questions in a Google Forms survey, describing their familiarity with dimension reduction and clustering algorithms and answering two questions about exploratory data analysis.  %All participants reported some degree of familiarity with clustering algorithms, but three of the participants reported no exposure to dimension reduction algorithms.

Following this, they were provided with a short description of the tasks they must complete and the goals of the study, after which they saw the dataset for the first time.  In the task (referred to as Organization Task), participants were asked to organize the observations in any way that they wished, though they were limited to grouping and spatialization operations (i.e., ``place two observations in the same group'' and ``place two observations some distance apart'').  Participants were instructed to think aloud in order to better capture their organizational process~\cite{nielsen2002getting,young2005direct}.
%After completing this organization, the second task (referred to as Update Task) asked them to update their organization given new information that two of the dimensions were more important than the other three.  These two dimensions were intentionally chosen as the dimensions that the participant used least in the Organization Task, in order to require a significant organizational update.  Additionally, the participants who received the abstract dataset were informed of the animal mapping and asked to comment on their organizational structure given the addition of labels.  
In addition to notes written in real-time by the proctor, this portion of the session was video recorded for later review.  Finally, participants completed a second survey of open-ended questions addressing their thoughts related to their personal analysis process.

% \reviewer{R1:  I finished the paper and was wishing that the experimenters would have simply asked, ``What would have helped you?'' to the participants to extract more input and ideas about the implications for design of computational tools.}

\subsection{Research Questions}

% \john{From rebuttal:  ``We plan to reorganize the RQs into groups as follows (noting the current RQ numbers, which will change):
% \begin{itemize}[noitemsep,nolistsep]
%     \item Participant analysis process (1,5)
%     \item Representations created (2,6)
%     \item Interactions with those representations (3, 4)
%     \item The effect of domain knowledge on the above (7)''
% \end{itemize}
% }

As previously noted, our overarching research question is to study the organizational processes of analysts when approaching a sensemaking task with an unfamiliar dataset.  We divide our more specific research questions into four broad themes, which correspond to the subsections of the Results that follow in the next section:

\begin{enumerate}[noitemsep,nolistsep]

    \item Participant Analysis Process
    \begin{enumerate}[noitemsep,nolistsep,label=(\Alph*)]
        \item How do analysts begin to evaluate an unfamiliar dataset, and what actions do they take during this initial evaluation?
        \item What are the overarching analysis strategies of study participants throughout the complete process of transforming an unfamiliar dataset into an organized space?
    \end{enumerate}
    
    \item Representations Created
    \begin{enumerate}[noitemsep,nolistsep,label=(\Alph*)]
        \item What types of grouping and spatial structures are created during the full analysis process?
        \item How do the individual dimensions appear within the organizational spaces created by participants?
    \end{enumerate}
    
    \item Interactions with Representations
    \begin{enumerate}[noitemsep,nolistsep,label=(\Alph*)]
        \item What types of grouping and spatial operations are performed during the full analysis process?
        \item How do participants approach individual dimensions vs collections of dimensions when exploring the dataset?
    \end{enumerate}
    
    \item The Effect of Domain Knowledge
    \begin{enumerate}[noitemsep,nolistsep,label=(\Alph*)]
        \item What is the role of external information in the analysis process? 
        \item How do the layouts created by the abstract condition participants change when they are provided with labeled information?
    \end{enumerate}
        
\end{enumerate}

\section{Results}

Using notes taken during the session and video recordings, we reviewed the actions of each participant throughout their analysis process.  During this review, we were purposefully searching for and noting the frequency of some behaviors, such as the types of cluster interactions previously noted in Section~\ref{sec:dr_clust}.  For the broader participant strategies that are described in this section, we used an open coding approach to describe these events, which we later synthesized into general themes.

\subsection{Participant Analysis Process}

In this section, we discuss the analysis process undertaken by the study participants, with foci on both how their analysis began and how their overall strategy developed.

\subsubsection{Beginning the Analysis (RQ1A)}

% \todo{
% \begin{itemize}[noitemsep,nolistsep]
%     \item Abstract data participants lay out all data points to examine, labeled data participants flipped through cards one-by-one
%     \item Both sets of participants mostly formed groups to begin analysis, labeled group were more likely to do a mix, abstract were the only ones to do mostly spacing
% \end{itemize}
% }

There were two primary methods by which participants approached the Organization Task.  The most common strategy, the \textbf{Grid Method}, was to begin by laying out the full dataset on the table, often in a grid pattern, in order to inspect the full dataset simultaneously.  Slightly less often, we saw the \textbf{Stack Method}, in which participants kept the index cards in a stack, inspecting each sequentially and determining its optimal location in the partially-organized space.  Several participants began by inspecting the top few cards in the stack before turning to one of the two primary patterns.  One participant in the labeled condition looked through the full stack to survey only the names of the animals, organizing the cards in the stack before following the sequential pattern.

Interestingly, there was a divide between the common approaches in the two groups.  Seven of the eight participants who received the abstract dataset followed the Grid Method, while only two of the eight participants did so with the labeled dataset.  Conversely, six of the eight participants who received the labeled dataset followed the Stack Method, while only one of the abstract dataset participants did so.

\begin{figure}[!tb]
    \centering
    \includegraphics[width=\linewidth]{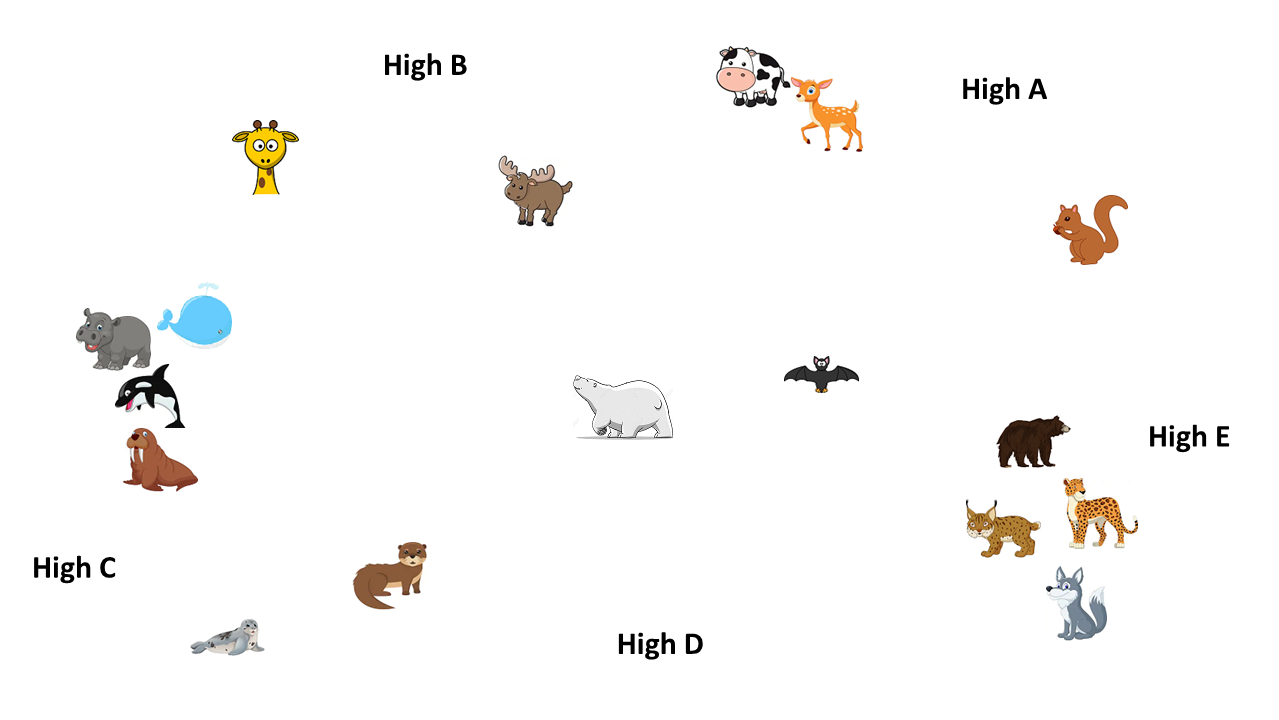}
    \caption{The radial layout created by Participant A2, in which each of the animals is drawn towards its highest attribute with additional effects by the other large attribute values.  This figure (and the other animal layout figures within this work) are recreations of the index card layouts created by the study participants.}
    \label{fig:A2}
\end{figure}

Only one of the participants in the abstract data condition primarily used spatialization actions to begin exploration the data.  Using the Stack Method, this participant created a radial layout in which observations were drawn towards five points that corresponded to full values of each of the dimensions (Fig.~\ref{fig:A2}), in a manner reminiscent of the Dust \& Magnet system~\cite{yi2005dust} where dust observations were drawn towards radially-positioned magnet points of high attribute value.  The remainder of the abstract dataset participants performed mostly grouping operations to explore the dataset, as did three of the labeled dataset participants.  The remaining five labeled condition participants performed a blend of grouping and spatialization operations, to the degree that neither category could be clearly considered a majority.

\subsubsection{Overarching Strategies (RQ1B)} \label{sec:study_strategies}

% \reviewer{R1:  In a couple places in the results section, I felt that the authors could leverage more related, prior research. For example, when discussing strategies that participants followed for performing the task, this reminded me of Kang's user studies of how people performed sensemaking tasks with (and without) the Jigsaw system. [IEEE TVCG 2011, Vol. 17, No. 5]~\cite{kang2011how} She found strategies liked examining each document first, picking one and exploring it in detail, etc. Some of these feel akin to the strategies this study uncovered. A second example happened when reading about the participant who brought in Canada as a factor. It is clear, as the article notes, that background knowledge and domain knowledge matter here. This led me to think of Peck's Best Paper from CHI '19~\cite{peck2019data} that discusses how all data is local and contextual. These two things aren't glaring omissions from the paper -- I simple feel that the authors could leverage connections to them and other papers to fortify their arguments and discussion.}

\begin{figure}[!tb]
    \centering
    \includegraphics[width=\linewidth]{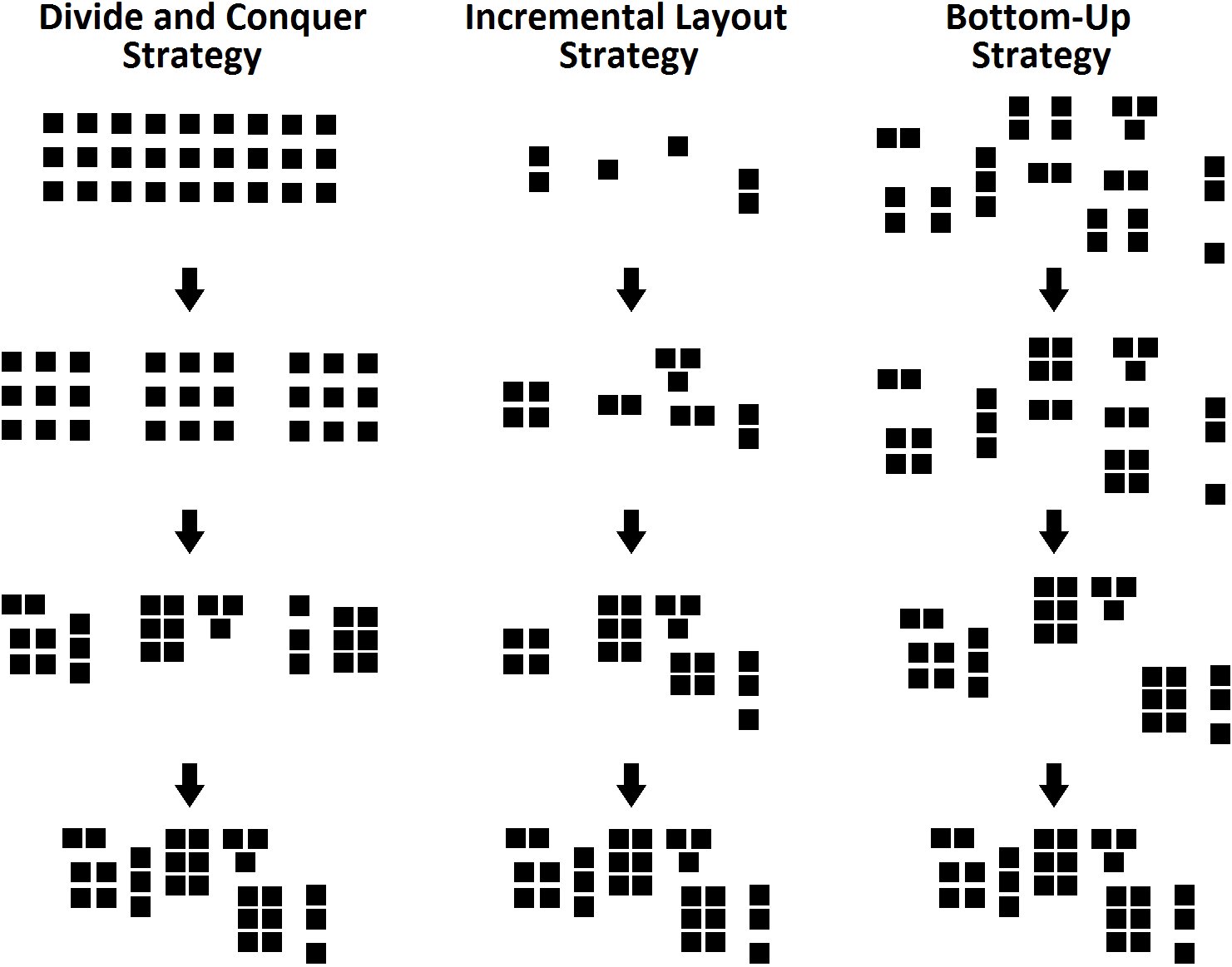}
    \caption{Representations of the three overarching strategies that participants used to organize their sensemaking space.  In these representations, the Divide and Conquer Strategy begins with the Grid Method, and the Incremental Layout Strategy begins with the Stack Method.}
    \label{fig:strategies}
\end{figure}

At a high level, we observed a tight coupling between spatialization and grouping actions performed by analysts.  Rather than adopting a purely group-first or layout-first mentality, participants in this study switched between the two frequently.  This was even true in cases where grouping or spatialization actions accounted for a great majority of the overall interaction total.  Further, we note that this complex relationship develops over time, where spatializations are used to drive grouping and grouping is used to drive spatializations.  This results in complex organizational spaces that were produced by the participants in this study.  We delve into these issues further in this section.

\begin{figure}[!b]
    \centering
    \includegraphics[width=\linewidth]{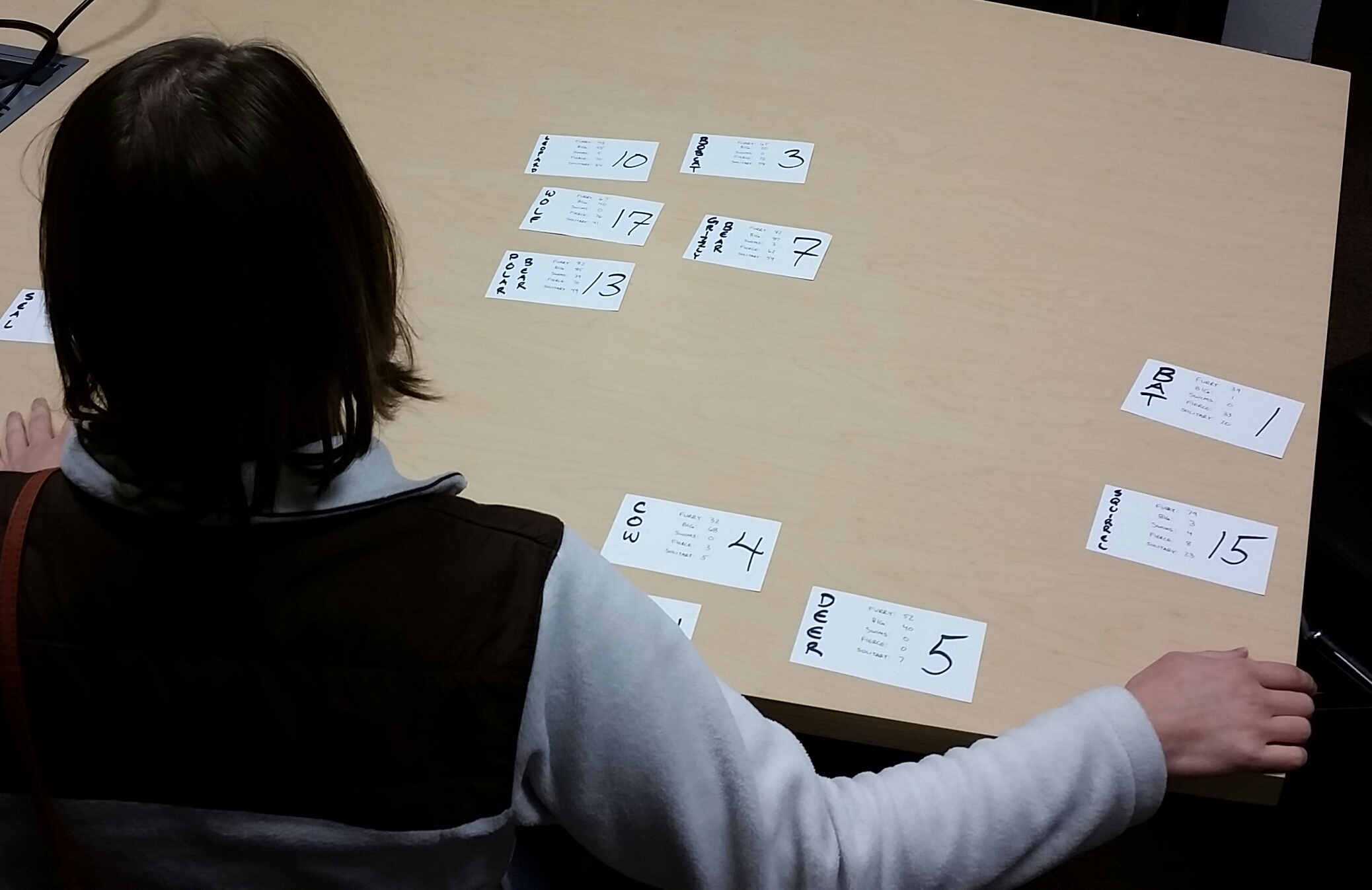}
    \caption{A participant from the labeled condition midway through her organizational process.  Groups of animals are apparent, and she is beginning to create spatial structures within the groups, following the Divide and Conquer strategy.}
    \label{fig:participant}
\end{figure}

There were three main strategies that participants performed when approaching the task (see Fig.~\ref{fig:strategies}), which we noted share some similarities to the investigative strategies observed by Kang et al.~\cite{kang2011how} in their document-based study using Jigsaw.  Though the switch from documents to quantitative data made the precise interactions and motivations different, the strategies that we note here show patterns that are similar.  For example, the Divide and Conquer strategy that we describe next resembles their Overview, Filter, and Detail strategy, while our Bottom-Up strategy is comparable to their Build from Detail.

The most common strategy seen is the \textbf{Divide and Conquer Strategy}, demonstrated in the photo in Fig.~\ref{fig:participant}.  In the first pass through the data, participants selected a single dimension and separate the observations into a small number of groups.  They then attempted to find meaning within these smaller groups, either by selecting another dimension to separate by or spatializing within the group.  After structuring the individual groups, they turned their attention to the full space and attempt to organize the large groups, with occasional refinement within the groups.

\begin{figure}[!tb]
    \centering
    \includegraphics[width=0.67\linewidth]{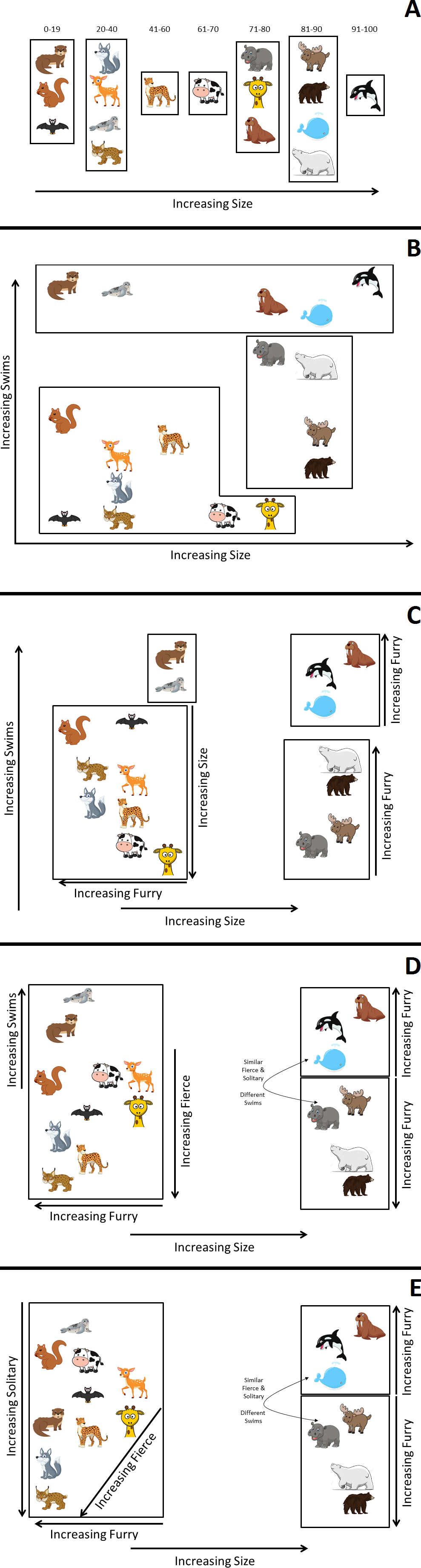}
    \caption{Five stages from the analysis produced by Participant L4.  The participant used the Divide and Conquer strategy, dividing the dataset into groups to be individually analyzed and refined.}
    \label{fig:development}
\end{figure}

An example of this strategy is provided by Participant L4 in Fig.~\ref{fig:development}.  In Panel~A at 4:47, she has binned the animals by size, creating seven temporary groups that increase in size from left to right across groups and from bottom to top within groups.  Panel~B at 9:04 includes a dimension for Swims, forming a structure that approximates a scatter plot.  At this point, she identifies three main groups:  swimming animals, sort-of swimming animals, and non-swimming animals.  In Panel~C at 16:46, she has decided to separate the swimming animals group as she began to sort within the groups by the Furry dimension.  The global Swims dimension still persists from bottom to top, but the global Size dimension is now discrete across the two columns of groups.  A local Size dimension was maintained vertically in the non-swimming animals group.  The Furry dimension was vertical in the larger swimming and sort-of swimming groups, but was horizontal in the non-swimming group (and was not clearly specified in the small swimming group).  In Panel~D at 21:03, the two groups of large animals were positioned closer together (though kept as separate groups) and the sort-of swimming animals group was flipped vertically because the Blue Whale, Moose, and Hippopotamus have similar Fierce and Solitary attributes.  The vertical axis of the smaller animals group (joined into a single group) now has a Fierce dimension with the non-swimming animals, though the Swims dimension is still maintained at the top of the group.  Finally in Panel~E at 31:21, the Fierce axis was rotated within the small animals group, and the vertical axis has been replaced with a Solitary dimension. 

The second most common strategy was the \textbf{Incremental Layout Strategy}.  This strategy was almost exclusive to the labeled data condition (A2 once again being the exception).  Participants considered each observation one at a time, adding them to a continually growing organizational space in the location which appeared most sensible.  Each of these additions was often a grouping operation, but could also be a spatialization operation in some cases.  Updates to the position of observations already positioned did occur, but were infrequent.  As the participants continued to add data, the physical size of the utilized space increased.  After all observations were added to the space, the participants began a more thorough refinement process.

A third strategy, not as common as the first two but still implemented by multiple participants, was the \textbf{Bottom-Up Strategy}.  Participants began similarly to Divide and Conquer, laying out all of the observations to view simultaneously.  However, their next step was to begin to build small groups of two or three similar observations, usually by only looking at a single dimension at first but then considering others.  After many of the observations had been placed in small groups, spatial relationships were created between the groups, often leading to the formation of larger groups.

%\todo{a figure showing timestamps and G/S operations performed by one of the analysts}

Regardless of the approach strategy taken by participants, they all followed an incremental pattern that consisted of a period of organization followed by a period of reflection, after which the process repeated.  Such incremental formalism has been demonstrated in previous studies and system use cases~\cite{bradel2014multimodel,endert2012semantic,shipman1999formality,kang2011how}, but took on a different form in this study.  As noted previously and confirmed by the post-survey, participants almost universally approached the organization by considering a single dimension at a time.  In doing so, they created an organizational space for one dimension, and then take a step back to consider the space.  They then proceeded to a second dimension, introducing its effects into the space gradually by individual animal or group, and then examined the global changes made to the structure.  This alternating pattern of sensemaking and synthesis segments connected the participants' local interactions to their global understanding of their organizational space. %\todo{support this with data, maybe a timeline figure}

\subsection{Representations Created}

In this section, we examine the grouping and spatial structures that were created by the participants during the course of their exploration.

\subsubsection{Grouping and Spatial Structures (RQ2A)} \label{sec:structures}

% \todo{
% \begin{itemize}[noitemsep,nolistsep]
%     \item Both participant sets were equally likely to create hierarchies and cross-cutting clusters
%     \item Axes mattered for participants in both groups
%     \item Both sets equally likely to identify outliers
%     \item Cluster sizes varied in both sets
%     \item Both sets equally likely to create internal cluster structures
% \end{itemize}
% }

\begin{figure}[!tb]
    \centering
    \includegraphics[width=\linewidth]{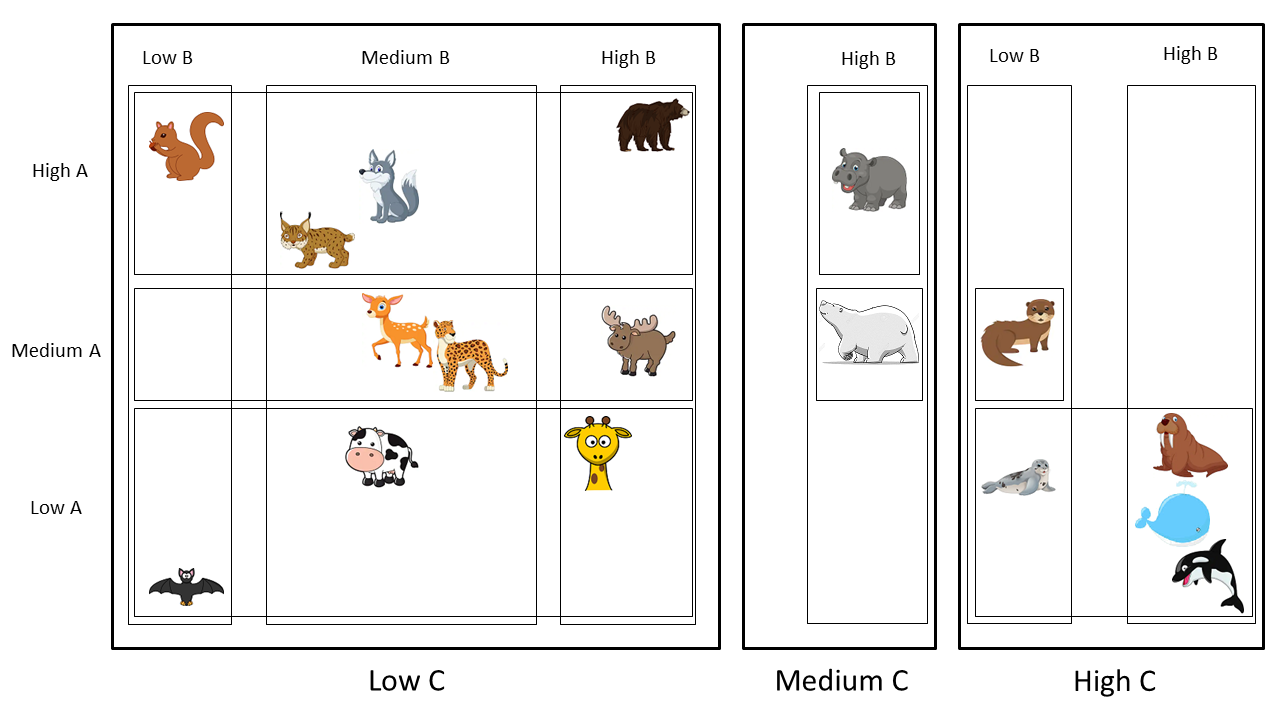}
    \caption{The complex group cross-cutting created by Participant A3.  This participant judged groups of abstract data by considering low, medium, and high values of dimensions A, B, and C (corresponding to Furry, Big, and Swims).  As a result, the axes of the organization were important to the group determinations.}
    \label{fig:A3}
\end{figure}

Both groups of participants were approximately equally likely to create hierarchies and cross-cutting groups in their organizational structures (4/8 labeled and 5/8 abstract).  Many participants did create internal spatial structures within their groups, but only a small subset clearly delineated groups within groups or groups overlapping groups (for example, Participant A3 in Fig.~\ref{fig:A3}).  In many cases, these internal groups were formed by breaking up a larger supergroup, though occasionally two subgroups were joined together to form children of a larger parent group.  These overlapping groups occasionally represented ``fuzzy'' or ``soft'' cluster assignments~\cite{zadeh1965fuzzy}.

\begin{figure}[!b]
    \centering
    \includegraphics[width=\linewidth]{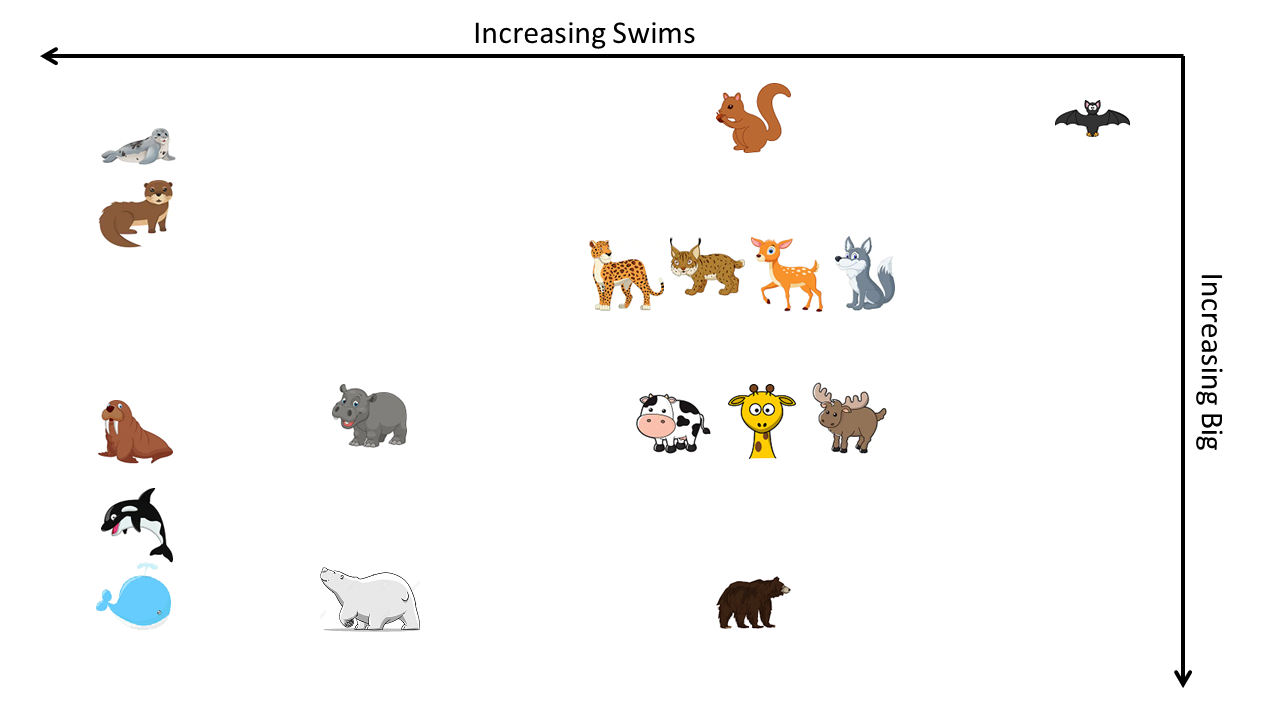}
    \caption{Participant L5 created a structure in which the axes were clearly an important feature.  Groups were refined based on the remaining three dimensions, but the majority of the structure was governed by the Swims and Big dimensions with which she began her analysis.  Attribute bins from the analysis of the Swims dimension are clearly still visible on the x-axis.}
    \label{fig:L5}
\end{figure}

Similarly, both groups of participants were equally likely to create organizational structures in which the axes mapped to dimensions in the data.  This is contrary to the properties of many dimension reduction algorithms, in which the axes have no direct mapping to the source data.  Often, such constructions resulted from participants' behavior in focusing on a single dimension at a time and organizing the observations on a spectrum along one or more dimensions (see Fig.~\ref{fig:L5}).  As axes were seen to be important, tools should clearly communicate the meaning of axes in a projection, potentially drawing inspiration from techniques seen in InterAxis~\cite{kim2016interaxis} and AxiSketcher~\cite{kwon2017axisketcher}.

We also noted that both groups of participants identified outliers in the dataset that they were hesitant to place in any group.  Near the beginning of their analysis, they referred to single-observation groups as groups (or the seeds of a group), but as they continued to structure observations in the space, they were more likely to refer to these observations as not fitting well with the others.  The participants who created cross-cutting groups often had subgroups with just a single observation in their organizational structure, but they were clear to identify those as equally belonging to two or more of the broader groups (again see Fig.~\ref{fig:A3}).

% The sizes of groups that the participants created was also quite broad.  All but one of the participants had at least one group that only contained a single observation, and most of the groups were small in cardinality.  However, some participants did occasionally create large groups encompassing 1/3 to 1/2 of the overall dataset.

Spatial meaning internal to groups was also seen by both participant conditions.  Often, the spatializations within the groups were designed to show differentiation within observations in the group (e.g., a size trend across the group), though occasionally the goal of the participants was to show relationships between members of the group and other parts of the space (e.g., an observation within the group that is quite similar to those in other groups).  As a consequence of the second, participants in both groups were equally likely to create global spatial structures that spanned the entire structure or governed large portions of their organization.

\begin{figure*}[!tb]
    \centering
    \includegraphics[width=0.95\linewidth]{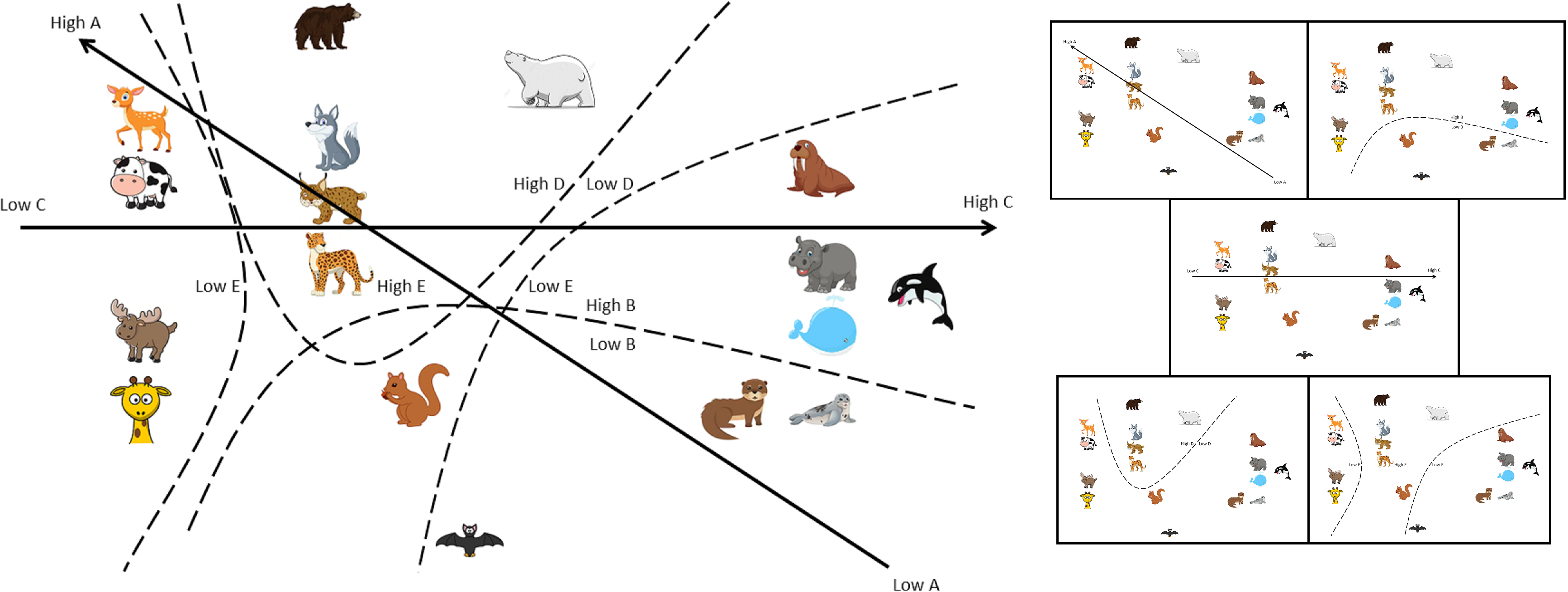}
    \caption{Participant A4 created a complex space, creating spectra (solid line arrows) for dimensions A and C (Furry and Swims) and distinct regions of high influence (dashed arcs) for dimensions B, D, and E (Big, Fierce, and Solitary).  The smaller subfigures to the right of the main space display the effect of each individual dimension on the overall projection with less clutter.}
    \label{fig:complex_spaces}
\end{figure*}

% \reviewer{R1:  In places, this feels a little low-level and can be just a little tedious, but clearly it is important. I like it best when the article is summarizing findings and using specific examples to briefly explain those generalizations, as opposed to the more detailed recaps. This sentiment builds up a little as one gets to the later discussions on RQ 6 and 7. There are a few places where being more specific would be helpful. For example, in the text at the bottom of the first column and top of second column on page 9, I was wondering which of the two groups the people with those sentiments came from. On page 10, the end of the first paragraph makes a jump to discussing parallels to t-SNE, subspace clustering, etc. That's a jarring and big jump. What if readers aren't familiar with those things? This simply needs more explanation. (This is in the table above as well.)}

\subsubsection{Complex Spaces (RQ2B)} \label{sec:complex_spaces}

After considering several dimensions, participants began to create complex spaces.  This was already seen through the hierarchical, cross-cutting set of groups created by Participant A3 (Fig.~\ref{fig:A3}).  Another example is seen within the structure created by Participant A4 (Fig.~\ref{fig:complex_spaces}), in which the participant created \textbf{spectra} for two of the dimensions and \textbf{influence regions} for the three remaining dimensions.  The spectra were not orthogonal, though the C dimension was aligned with the x-axis. These are represented by the solid arrows in Fig.~\ref{fig:complex_spaces}. The three regions likewise overlapped in some places but not others, indicating portions of the space in which one dimension had a great deal of influence in determining how the participant structured the layout and groups.  These are represented by the dashed arcs in the same figure.

This runs counter to the common method of creating dimensionally-reduced projections, in which the entire space is governed by a single weight vector (for example,~\cite{bradel2014multimodel,dowling2018sirius,self2018observation,wenskovitch2017observationlevel}).  Creating clusters that contain independent, internal weight vectors, as well as maintaining a global space, presents one solution to this challenge.  For the example presented by A4, the low~B region could be defined as a cluster that still maintains the influence of low~A and high~C.  This finding presents opportunities for the introduction of subspace clustering techniques to support such complex spaces, as seen in previous works analyzing high-dimensional data~\cite{parsons2004subspace,kriegel2009clustering}.

\begin{figure}[!b]
    \centering
    \includegraphics[width=0.95\linewidth]{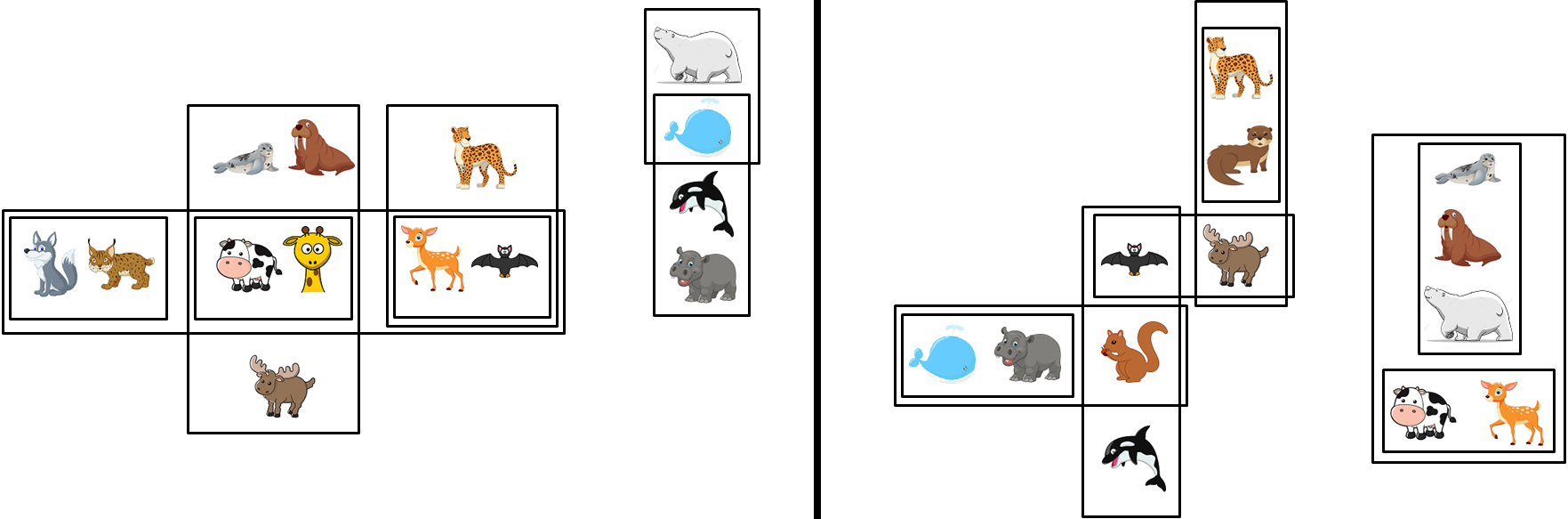}
    \caption{Participant A6 created complex hierarchical and cross-cutting group structures during her analysis.}
    \label{fig:A6}
\end{figure}

Further, these complex spaces are not limited to areas of attribute influence.  Participants also created complex structures of hierarchical, cross-cutting groups.  For example, Participant A6 created several complex sets of groups at various points in her analysis, two of which are provided in Fig.~\ref{fig:A6}.  This participant's bottom-up process of connecting smaller clusters into larger groupings also reflects Gillam's connectedness findings for cognitive grouping~\cite{gillam2001varieties}.

\subsection{Interactions with Representations}

Having established the created structures, we now examine the operations performed on these structures during the course of analysis.

\subsubsection{Grouping and Spatial Operations (RQ3A)} \label{sec:operations}

% \todo{
% \begin{itemize}[noitemsep,nolistsep]
%     \item Most participants started off with large groups based on one dimension and then subdivided
%     \item Labeled data participants performed a broader variety of cluster operations (more likely to create, join, remove)
% \end{itemize}
% }

\begin{figure}[!b]
    \centering
    \includegraphics[width=0.8\linewidth]{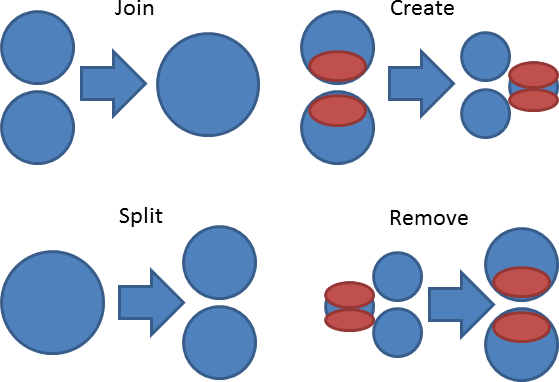}
    \caption{Grouping operations in the participant organizational strategies.  Group join operations destroy two original groups to create a new group, group split operations destroy one original group to create two new groups, group creation operations take a portion of one or more groups to create a new group (while leaving the originals), and group remove operations destroy one original group and distribute its members to one or more existing groups.}
    \label{fig:cluster_operations}
\end{figure}

We recorded instances in which participants performed four different types of grouping operations:  create, remove, join, and split.  These are differentiated in Fig.~\ref{fig:cluster_operations}.  The most common method by which participants in both groups approached the organizational task was to start with large groups and then subdivide.  As a result, splitting an existing group into smaller groups was the most common grouping operation performed.  A majority of participants also joined groups together at some point in their analysis, usually when considering a dimension for the first time and noticing new similarities among the observations.  Only two participants in the abstract condition (A2 and A6) performed operations to create an entirely new group from observations previously in several other groups.  None of the participants in the abstract condition removed a group and allocated its members into several other groups.  In contrast, five of the eight participants in the labeled condition created and three of the eight removed a group.

Spatial operations were occasionally the driving force behind group creation.  Participants in both the labeled and abstract conditions often identified collections of observations that were similar and positioned them close together spatially before identifying that collection as a group.  Conversely, groups were often used to drive spatial operations as well, especially when participants were refining group memberships and identifying distances within groups.

Participants also reported that distances within groups were more important to their structure than distances between groups.  This was partially a result of the cognitive difficulty in mentally computing a distance between groups of observations as opposed to computing a distance between a pair of observations.  More meaningfully, participants reported that these fine-grained differences between observations within a group were more relevant to their understanding of group structure than were the differences between the groups themselves.  In other words, it was enough for participants to say ``these groups are different,'' but they felt the need to incorporate more spatial detail when saying ``this is why these observations within a group are different.''

\subsubsection{Decision Making (RQ3B)} \label{sec:decisionmaking}

% \todo{
% \begin{itemize}[noitemsep,nolistsep]
%     \item All but one participant examined one dimension at a time
%     \item Many binary decisions in both groups
%     \item All but one participant performed attribute binning
%     \item 3 participants (2A, 1L) created features
%     \item 1 participant (L) ignored all dimensions and focused on names
% \end{itemize}
% }

With the exception of A2, all participants in both groups spent the majority of their analysis considering only a single dimension at a time, confirming the observation seen in previous studies that analysts struggle to think high-dimensionally~\cite{andrews2012analysts,self2018observation}.  Additionally, participants frequently processed attributes by either making binary decisions (e.g., divide observations by an attribute value greater than or less than~50), or alternatively by creating a small number of bins to discretely group observations by a single dimension.  The participants commented that both the binary decisions and the binning operations were intentionally made so that they could focus their attention on subsets of the observations rather than the entire collection.

Two of the participants in the abstract condition created features from combinations of provided features while exploring the data, in both cases to reduce the amount of information that they were trying to cognitively process.  Participant A5 computed the median of all five dimensions in other to perform an initial grouping, while A4 computed the difference between the final two variables she had not yet considered.  Two of the participants from the labeled condition also created features, but these came from domain knowledge of the labeled animals instead.  Participant L5 introduced both flying and speed features into her layout, while the other created groups that incorporated the likelihood of finding these animals in a zoo.  This last participant, L1, also reported that she focused primarily on the animal labels, and did not consider the attribute values until making final refinements to the space.  Kopanas et al. found similar data transformations and feature interpretations in their data mining study~\cite{kopanas2002role}. 

\subsection{The Effect of Domain Knowledge}

We now look at the differences between the labeled and abstract groups, with a particular focus on the role of external information.% plays in the exploration of the participants.

\subsubsection{External Knowledge (RQ4A)} \label{sec:external}

Participants who received the labeled dataset often made use of their external knowledge about the animals that was not contained on the cards.  These behaviors in response to external and domain knowledge reinforce the ``Data is Personal'' work by Peck et al.~\cite{peck2019data}, particularly that personal experience drives decisions.  Similar experience-driven behaviors have been seen in the cognitive psychology field, as evidenced by the experiments of Zemel et al~\cite{zemel2002experience}.

External knowledge influenced their organizational spaces in three ways.  First, introducing external knowledge allowed the participants to begin forming groups of animals prior to closely examining the data.  For example, Participant L1 created a number of groups without even considering the data, including features corresponding to environment, diet, and probability of locating the animal in Canada.  Likewise, Participant L5 introduced a flight group for the Bat into her organization, keeping it separate from all other animals as a result without considering the attribute values on the card.  

Second, external knowledge was used to structure and spatialize within groups.  For example, Participant L5 introduced a speed dimension into her organization, which permitted her to break up the land-dwelling (and non-flying) animals into groups based on how quickly they moved.  The final result of this introduction was a group that contained the Deer along with the Wolf, Leopard, and Bobcat.  Occasionally, incorrect domain knowledge could also affect their organization, as seen when Participant L3 used external knowledge about the diet of animals to create transition groups.  For example, the Hippopotamus was used as a transition animal between Aquatic and non-Aquatic animals, not because of its mid-range Swims score, but because of the fact that she felt this animal was likely to consume fish while spending time in water as well as plants when spending time on land.

Third, external knowledge about the animals was used as confirmation, checking to see if the groups the participants created were sensible after performing a dimension-specific organizational step.  Participant L2, for example, frequently questioned his positioning of the Bat throughout his iterative organizational steps.  Despite the attributes on the cards supporting his decisions, he continued to second-guess its position and occasionally to modify it based on his external knowledge.  Participant L4 also reported scanning the names of the animals in the groups regularly, searching for refinements that could be made to the structure internal to groups based solely on those labels.

\subsubsection{``The Big Reveal'' (RQ4B)}

Participants who received the abstract dataset were informed of its animal dataset source after completing their organizational space.  The animals cards were overlaid above the abstract cards, and participants were asked to comment on the structure and relationships that were now apparent after seeing the labeled data, effectively replicating the confirmation use from the previous subsection.  

In general, participants reported being satisfied with the layout that they created in the abstract data.  Many had an Aquatic group of animals due to the high-C score, and they quickly picked up on this relationship.  Similarly, participants who prioritized the D dimension saw groups of predators in their organization.  The Bat often was a frustration point for participants, expressing dissatisfaction with its position in the space similar to the reaction seen by participant L2.  

When asked what they would change in their space given this new information, most participants reported that they would increase the density of the groups in their space, moving groups of similar animals closer together after understanding their relationship to each other.  This was often seen in any Aquatic groups, Predator groups, and among the Whales.  Interesting, this behavior was not necessarily true with the Bears; participants were more willing to keep these animals separate and to occasionally increase their separation despite the common genus, often reporting that this was counter to their initial reaction when noting the existence of two bears.

The incorporation of external knowledge by the labeled condition participants, and the changes made to the layouts of the abstract condition participants once the labeled information was provided, both support the utility of semantic interaction in interactive systems.  The goal of semantic interaction~\cite{endert2012semantic} is to permit the analyst to focus their exploration on the observations themselves, rather than carefully examining data values and fine-tuning the parameters of underlying models.  Because the labeled group participants often brought in external knowledge, they were able to add additional detail to their organizational structures that was not contained on the index cards.  The updates made by the abstract condition participants after the presentation of labels show that their spaces would have been differently structured if they had access to this information throughout the study.

% \todo{confirmation of semantic interaction?  bias towards thinking animals are more similar than they really are?  difficulty with managing numbers and delineating clusters?}

\section{Discussion}

The results collected from this study have yielded valuable information towards how humans approach exploring and organizing data.  Such information can be used to guide the design of interactive visual analytics systems in the future.

\subsection{Post-Survey} \label{sec:postsurvey}

All participants provided responses to the survey that followed this study.  These questions addressed their interpretation of their strategy, easy and difficult parts of the analysis, and their thoughts on the usefulness and meaningfulness of grouping and spatialization actions.

\textbf{Approaching the Task:} Participants reported approaching the task by considering a single dimension at a time, confirming observations from the study.  The difficulty that the participants experienced when attempting to think high-dimensionally suggests the need for computational support in similar organizational tasks.  Each dimension was selected by searching for features in the dataset that seemed to be most useful or representative, either due to the overall distribution, outliers, or common values.  When considering each dimension, participants sought out commonalities between the observations, building large groups or binning the observations in a spectrum and refining the bins.  Participants who received the labeled set also used external knowledge of the animals to create and organize groups.  

\textbf{Number of Interactions:} When asked whether they thought they performed more grouping or spatialization interactions, participants gave a variety of responses.  Among those who believed they performed more grouping operations, they noted that their overarching strategy was to isolate groups within the data in order to make future processing simpler.  Those who believed that they performed more spatialization interactions generally reported that they were careful when refining the organization in later stages, leading to that majority.  Some participants reported that they couldn't determine which class was the majority.

\begin{table*}[!tb]
  \renewcommand{\arraystretch}{1.2}
  \newcolumntype{M}{>{\arraybackslash} m{11cm}} %Centers vertically 
  %if using array.sty, it might be a good idea to tweak the value of
  %\extrarowheight as needed to properly center the text within the cells
  \caption{A summary of the main findings uncovered by this study and corresponding design recommendations for interactive data exploration tools.}
  \label{tbl:findings}
  \centering
  \resizebox{\linewidth}{!}{
  \begin{tabular}{ M | M }
    \textbf{Finding}  &  \textbf{Design Recommendation} \\
    \hline
    A tight coupling was seen between grouping and spatialization actions, frequently switching between these operations.  Participants formed groups in order to make future spatializations easier, and also formed spatializations to make future groupings either.  &  Systems should be designed so that analysts can perform such operations at any time during the analysis process, as opposed to creating sequential phases.  Algorithms should be selected or designed to support these highly-incremental feedback loops. \\
    \hline
    The axes often had meaning in the organizational spaces created by participants, with one or more dimensions frequently parallel or orthogonal to the front of the table.  &  Aligning an important property (e.g., the first principal component or the dimension with the highest weight) to an axis can boost the interpretability of the projection.  Alternatively, the meaning of the axes should be clearly communicated. \\
    \hline
    A common trend was to progress from large groups to small groups, splitting groups rather than joining them.  &  While many types of cluster operations (e.g., joining, creating, removing) should be supported, an efficient interaction for splitting clusters should be a design priority. \\
    \hline
    Participants reported that distances within groups were more important to their structure than distances between groups.  &  Algorithms that favor local structures (e.g., t-SNE, subspace clustering) may be better representations of an analyst's understanding of data than those which do not. \\    
    \hline
    To further reduce the complexity of the data, participants often binned the observations into smaller groups or separated the observations with a binary decision.  &  Potential means of designing such an efficient cluster-splitting interaction include automatically binning observations or separating them by an analyst-specified threshold. \\
    \hline
    Participants primarily explored one dimension at a time, expressing frustration with trying to consider all dimensions at once.  &  Computational support is key for efficiently communicating multidimensional information to the analyst. \\
    \hline
    Participants often created complex spaces that combined dimension spectra with dimension regions of influence.  &  Tools should support complex spaces (e.g., subspace clustering and other techniques that favor local structures) rather than using a single weight vector to express the full space. \\ %Interactions should be provided to support both the analyst and system in creating these layouts. \\
    \hline
    Participants in the labeled group often brought their external knowledge into their organizational structure, adding additional animal properties that were not provided.  &  Systems should provide interactions for annotation or other notes, permitting analysts to inject information into the system. \\
	\hline
  \end{tabular}
  }
\end{table*}

\textbf{Task Difficulty:} Participants generally reported that the easiest part of the Organization Task was the beginning or the end of the analysis.  Some reported that selecting a starting point, picking the initial groups, or creating the initial spatial structure was easiest.  Others noted that making final refinements within and between groups at the end of the analysis was easiest.  Most participants reported the mid-stages of organization to be the most challenging, needing to balance updates to existing groupings of data when examining a new dimension with maintaining existing spatial relationships.  Participant A1 did report that the amount of data seen after laying out all of the cards initially was overwhelming, but that did not stop her from arbitrarily selecting a starting dimension for analysis.

\textbf{Operation Usefulness:} Participants were also evenly split between considering the grouping or the spatialization actions to be more useful or more meaningful.  Those who felt more positively about the groupings mentioned summarizing the big picture and making sense of the overall space visually, while those who felt more positively about the spatializations thought that these actions made them more careful in their analysis.  Both groups also mentioned that their operation preference for this question impacted the other.  Those who preferred grouping actions noted that it made spatialization actions easier to perform, while those who preferred spatialization actions noted that it made the task of creating meaningful groups easier.

A related observation while running the study is the role of terminology, particularly with the grouping operation.  There were a number of times when participants were clearly separating the observations into piles, but they were somewhat hesitant to define this organization as a ``cluster'' or a ``group.'' Frequently, we found ourselves using a variety of terms when inquiring about the structures that the participants were creating (e.g., ``Do you consider this a group, or is it a cluster, or an organizational construct, or a bin, or a collection, or ...?'').  To the participants, the terms ``cluster'' and ``group'' had a different, deeper semantic meaning than a simple ``pile'' of observations.  In order to be classified as a ``cluster'' or ``group,'' participants often wanted to perform multiple iterations of analysis to ensure that more than just a single property defined the collection of observations.

\subsection{Recommendations for Tool Design} \label{sec:designrecs}

Table~\ref{tbl:findings} collects findings that were uncovered while running this study, each of which is described in further detail in the preceding sections.  Each of these findings is accompanied by a design recommendation for interactive visualization systems that support these interactions in the study domain (high-dimensional quantitative data).  Existing systems in this area (e.g., Castor~\cite{wenskovitch2017observationlevel} and Pollux~\cite{wenskovitch2019pollux}) begin to address some of these interaction and design issues, but do not fully address them all.  For example, neither system makes an effort to map an important dimension to an axis, support is not provided for efficient cluster splitting, and both use a single weight vector to express the full space.  Further development of these systems to reflect the lessons learned from this study will improve both analytical power and user experience or these and similar tools.

The overarching organizational strategies demonstrated by participants can also inform the choice and design of algorithms that support analysis in these systems.  For example, participants who used the Divide and Conquer strategy created spaces that differed between groups as of result of their separate analysis within each group.  This suggests the benefit of algorithms that favor local structures, such as t-SNE~\cite{maaten2008visualizing} for projection layout and subspace clustering~\cite{tatu2012subspace,mueller2009evaluating} for grouping.

It is important to note that the goal of including such algorithms is to better enable this organizational process, not to compute the final results.  There is great benefit to the exploration processes of the analysts, as they continue to refine hypotheses and structure observations.  An analyst who is presented with an organizational space created by the study participants will almost certainly walk away with conclusions that differ from the participant who created the space.

% \todo{map overarching strategies to algorithmic stuff.  e.g., divide and conquer strategy maps to localized feedback}

% \todo{algorithms should enable this process, not compute the final result}

\subsection{Limitations and Future Work}

The primary limitation of this work is that this study has only been tested on a single dataset rather than experimenting with datasets of various sizes (cardinality of observations and dimensions), types (documents), and levels of complexity (floating-point observations, conflicting dimensions, and confounding variables).  We recognize that the results found in this study and the recommendations presented in Table~\ref{tbl:findings} are limited in applicability to our domain of quantitative data.  For time-series or ensemble data, these recommendations may not apply (and the analysis undertaken by user will be different as well).  Additionally, this study was limited to investigating how an analyst thinks about and organizes information in a space that they create; however, we do not consider how an analyst will interpret an organizational space that is created for them by an analytical system.  Such a follow-up study is necessary to supplement the design recommendations from Sect.~\ref{sec:designrecs}. 

Future studies with other datasets can continue to explore this space.  In particular, running this study with documents could yield different results, as participants will likely focus on abstract conceptual dimensions rather than term frequency-based computational dimensions, as the dimensions are not obvious in the text.  Likewise, we expect that there will be additional focus on grouping operations at the beginning of the analysis process as the participants search for common topics or themes in the documents.  Comparing the results of this work to a document study represents a promising direction for future research.

This study yielded a number of additional questions worthy of investigation.  We plan to investigate the similarities between the final layouts of all participants by encoding groups and relative distances between observations into a high-dimensional dataset, and then visualize the result in a tool such as Andromeda~\cite{self2018observation} or SIRIUS~\cite{dowling2018sirius}.  These tools can also enable our understanding of which spatial or grouping properties cause such similarities between the spaces.  We can also perform similar analyses on the individual layouts, enabling us to understand bias in the layouts created by the participants.  A followup study using a computational system could present this bias information to the analyst in real time, in order to learn how participants react to that information.

% \chris{FW:  after looking at Alberto’s stuff, i’m realizing we need to do your study again with documents.  i think people will think very differently about it.  e.g. they wont focus on one or dimensions first, because the dimensions are not obvious in text.  so they will focus more on abstract conceptual dimensions, which makes it more interesting.  comparing the two studies would be interesting too.}

% \todo{how similar are these layouts?  how can we best measure this?}

% \todo{what would participants do with bias information?}

% \chris{might be interesting to put the users arrangements into Andromeda / Pollux and see what it learns. does andromeda confirm what participants said they were emphasizing in their layouts?}

\section{Conclusion}

In this work, we experiment with a labeled and abstract set of data to examine how analysts approach and organize an unfamiliar dataset.  We wish to understand the cognitive processes that underlie the approach that analysts take when trying to find insight in data.  We found that participants used groups to create spatial structures as well as spatial structures to form groups.  Participants created hierarchies and cross-cutting groups in their organizational structures, and frequently approached the task by creating large groups and subdividing them to refine additional structure.  The complex spaces created by participants hint towards structures that should be supported in interactive applications.  We summarize a list of main findings and corresponding system design recommendations in Table~\ref{tbl:findings}.

\acknowledgments{
The authors wish to thank those who participated in this study, as well as the reviewers for their insightful comments.}

\bibliographystyle{abbrv-doi}

\bibliography{sample}

\begin{thebibliography}{10}

\bibitem{amar2005lowlevel}
R.~{Amar}, J.~{Eagan}, and J.~{Stasko}.
\newblock Low-level components of analytic activity in information
  visualization.
\newblock In {\em IEEE Symposium on Information Visualization, 2005. INFOVIS
  2005.}, pp. 111--117, 2005.

\bibitem{andrews2010space}
C.~Andrews, A.~Endert, and C.~North.
\newblock Space to think: Large high-resolution displays for sensemaking.
\newblock In {\em Proceedings of the SIGCHI Conference on Human Factors in
  Computing Systems}, CHI '10, pp. 55--64. ACM, New York, NY, USA, 2010. doi:
  {{%
10\hspace{.1pt}\discretionary{.}{%
}{.}\hspace{.4pt}1145\discretionary{/}{%
}{/}1753326\hspace{.1pt}\discretionary{.}{%
}{.}\hspace{.4pt}1753336}}


\bibitem{andrews2012analysts}
C.~{Andrews} and C.~{North}.
\newblock Analyst's workspace: An embodied sensemaking environment for large,
  high-resolution displays.
\newblock In {\em 2012 IEEE Conference on Visual Analytics Science and
  Technology (VAST)}, pp. 123--131. IEEE, Oct 2012. doi: {{%
10\hspace{.1pt}\discretionary{.}{%
}{.}\hspace{.4pt}1109\discretionary{/}{%
}{/}VAST\hspace{.1pt}\discretionary{.}{%
}{.}\hspace{.4pt}2012\hspace{.1pt}\discretionary{.}{%
}{.}\hspace{.4pt}6400559}}


\bibitem{andrienko2009interactive}
G.~{Andrienko}, N.~{Andrienko}, S.~{Rinzivillo}, M.~{Nanni}, D.~{Pedreschi},
  and F.~{Giannotti}.
\newblock Interactive visual clustering of large collections of trajectories.
\newblock In {\em 2009 IEEE Symposium on Visual Analytics Science and
  Technology}, pp. 3--10, 2009.

\bibitem{barsalou1983ad}
L.~W. Barsalou.
\newblock Ad hoc categories.
\newblock {\em Memory \& cognition}, 11(3):211--227, 1983.

\bibitem{basu2010assisting}
S.~Basu, D.~Fisher, S.~M. Drucker, and H.~Lu.
\newblock Assisting users with clustering tasks by combining metric learning
  and classification.
\newblock In {\em Twenty-Fourth AAAI Conference on Artificial Intelligence},
  2010.

\bibitem{baylis1992visual}
G.~C. Baylis and J.~Driver.
\newblock Visual parsing and response competition: The effect of grouping
  factors.
\newblock {\em Perception \& Psychophysics}, 51(2):145--162, 1992.

\bibitem{boudjeloudassala2016interactive}
L.~Boudjeloud-Assala, P.~Pinheiro, A.~Blansché, T.~Tamisier, and B.~Otjacques.
\newblock Interactive and iterative visual clustering.
\newblock {\em Information Visualization}, 15(3):181--197, 2016. doi: {{%
10\hspace{.1pt}\discretionary{.}{%
}{.}\hspace{.4pt}1177\discretionary{/}{%
}{/}1473871615571951}}


\bibitem{bradel2014multimodel}
L.~Bradel, C.~North, L.~House, and S.~Leman.
\newblock Multi-model semantic interaction for text analytics.
\newblock In {\em 2014 IEEE Conference on Visual Analytics Science and
  Technology (VAST)}, pp. 163--172, Oct 2014. doi: {{%
10\hspace{.1pt}\discretionary{.}{%
}{.}\hspace{.4pt}1109\discretionary{/}{%
}{/}VAST\hspace{.1pt}\discretionary{.}{%
}{.}\hspace{.4pt}2014\hspace{.1pt}\discretionary{.}{%
}{.}\hspace{.4pt}7042492}}


\bibitem{brehmer2014interviews}
M.~Brehmer, M.~Sedlmair, S.~Ingram, and T.~Munzner.
\newblock Visualizing dimensionally-reduced data: Interviews with analysts and
  a characterization of task sequences.
\newblock In {\em Proceedings of the Fifth Workshop on Beyond Time and Errors:
  Novel Evaluation Methods for Visualization}, BELIV '14, pp. 1--8. ACM, New
  York, NY, USA, 2014. doi: {{%
10\hspace{.1pt}\discretionary{.}{%
}{.}\hspace{.4pt}1145\discretionary{/}{%
}{/}2669557\hspace{.1pt}\discretionary{.}{%
}{.}\hspace{.4pt}2669559}}


\bibitem{brown2012disfunction}
E.~T. {Brown}, J.~{Liu}, C.~E. {Brodley}, and R.~{Chang}.
\newblock Dis-function: Learning distance functions interactively.
\newblock In {\em 2012 IEEE Conference on Visual Analytics Science and
  Technology (VAST)}, pp. 83--92, 2012.

\bibitem{calinski1974dendrite}
T.~Cali{\'n}ski and J.~Harabasz.
\newblock A dendrite method for cluster analysis.
\newblock {\em Communications in Statistics-theory and Methods}, 3(1):1--27,
  1974.

\bibitem{choo2013utopian}
J.~{Choo}, C.~{Lee}, C.~K. {Reddy}, and H.~{Park}.
\newblock Utopian: User-driven topic modeling based on interactive nonnegative
  matrix factorization.
\newblock {\em IEEE Transactions on Visualization and Computer Graphics},
  19(12):1992--2001, 2013.

\bibitem{chuang2014human}
J.~Chuang and D.~J. Hsu.
\newblock Human-centered interactive clustering for data analysis.
\newblock In {\em Conference on Neural Information Processing Systems (NIPS).
  Workshop on Human-Propelled Machine Learning}, 2014.

\bibitem{chung2018savil}
H.~Chung and C.~North.
\newblock Savil: Cross-display visual links for sensemaking in display
  ecologies.
\newblock {\em Personal Ubiquitous Comput.}, 22(2):409--431, Apr. 2018. doi:
  {{%
10\hspace{.1pt}\discretionary{.}{%
}{.}\hspace{.4pt}1007\discretionary{/}{%
}{/}s00779\discretionary{%
}{-}{-}017\discretionary{%
}{-}{-}1091\discretionary{%
}{-}{-}4}}


\bibitem{chung2014visporter}
H.~Chung, C.~North, J.~Z. Self, S.~Chu, and F.~Quek.
\newblock Visporter: Facilitating information sharing for collaborative
  sensemaking on multiple displays.
\newblock {\em Personal Ubiquitous Comput.}, 18(5):1169--1186, June 2014. doi:
  {{%
10\hspace{.1pt}\discretionary{.}{%
}{.}\hspace{.4pt}1007\discretionary{/}{%
}{/}s00779\discretionary{%
}{-}{-}013\discretionary{%
}{-}{-}0727\discretionary{%
}{-}{-}2}}


\bibitem{cockburn2002evaluating}
A.~Cockburn and B.~McKenzie.
\newblock Evaluating the effectiveness of spatial memory in 2d and 3d physical
  and virtual environments.
\newblock In {\em Proceedings of the SIGCHI Conference on Human Factors in
  Computing Systems}, CHI '02, pp. 203--210. ACM, 2002. doi: {{%
10\hspace{.1pt}\discretionary{.}{%
}{.}\hspace{.4pt}1145\discretionary{/}{%
}{/}503376\hspace{.1pt}\discretionary{.}{%
}{.}\hspace{.4pt}503413}}


\bibitem{coden2017method}
A.~Coden, M.~Danilevsky, D.~Gruhl, L.~Kato, and M.~Nagarajan.
\newblock {\em A Method to Accelerate Human in the Loop Clustering}, pp.
  237--245.
\newblock SIAM, 2017. doi: {{%
10\hspace{.1pt}\discretionary{.}{%
}{.}\hspace{.4pt}1137\discretionary{/}{%
}{/}1\hspace{.1pt}\discretionary{.}{%
}{.}\hspace{.4pt}9781611974973\hspace{.1pt}\discretionary{.}{%
}{.}\hspace{.4pt}27}}


\bibitem{curiel1998mental}
J.~M. Curiel and G.~A. Radvansky.
\newblock Mental organization of maps.
\newblock {\em Journal of Experimental Psychology: Learning, Memory, and
  Cognition}, 24(1):202, 1998.

\bibitem{ding2007adaptive}
C.~Ding and T.~Li.
\newblock Adaptive dimension reduction using discriminant analysis and k-means
  clustering.
\newblock In {\em Proceedings of the 24th International Conference on Machine
  Learning}, ICML '07, pp. 521--528. ACM, New York, NY, USA, 2007. doi: {{%
10\hspace{.1pt}\discretionary{.}{%
}{.}\hspace{.4pt}1145\discretionary{/}{%
}{/}1273496\hspace{.1pt}\discretionary{.}{%
}{.}\hspace{.4pt}1273562}}


\bibitem{dobrynin2005sophia}
V.~{Dobrynin}, D.~{Patterson}, M.~{Galushka}, and N.~{Rooney}.
\newblock Sophia: an interactive cluster-based retrieval system for the ohsumed
  collection.
\newblock {\em IEEE Transactions on Information Technology in Biomedicine},
  9(2):256--265, 2005.

\bibitem{dourish1999building}
P.~Dourish, J.~Lamping, and T.~Rodden.
\newblock Building bridges: Customisation and mutual intelligibility in shared
  category management.
\newblock In {\em Proceedings of the International ACM SIGGROUP Conference on
  Supporting Group Work}, GROUP '99, pp. 11--20. ACM, New York, NY, USA, 1999.
  doi: {{%
10\hspace{.1pt}\discretionary{.}{%
}{.}\hspace{.4pt}1145\discretionary{/}{%
}{/}320297\hspace{.1pt}\discretionary{.}{%
}{.}\hspace{.4pt}320299}}


\bibitem{dowling2018sirius}
M.~Dowling, J.~Wenskovitch, J.~Fry, S.~Leman, L.~House, and C.~North.
\newblock Sirius: Dual, symmetric, interactive dimension reductions.
\newblock {\em IEEE Transactions on Visualization and Computer Graphics},
  25(1):172--182, Jan 2019. doi: {{%
10\hspace{.1pt}\discretionary{.}{%
}{.}\hspace{.4pt}1109\discretionary{/}{%
}{/}TVCG\hspace{.1pt}\discretionary{.}{%
}{.}\hspace{.4pt}2018\hspace{.1pt}\discretionary{.}{%
}{.}\hspace{.4pt}2865047}}


\bibitem{dowling2019interactive}
M.~Dowling, N.~Wycoff, B.~Mayer, J.~Wenskovitch, S.~Leman, L.~House, N.~Polys,
  C.~North, and P.~Hauck.
\newblock Interactive visual analytics for sensemaking with big text.
\newblock {\em Big Data Research}, 16:49 -- 58, 2019. doi: {{%
10\hspace{.1pt}\discretionary{.}{%
}{.}\hspace{.4pt}1016\discretionary{/}{%
}{/}j\hspace{.1pt}\discretionary{.}{%
}{.}\hspace{.4pt}bdr\hspace{.1pt}\discretionary{.}{%
}{.}\hspace{.4pt}2019\hspace{.1pt}\discretionary{.}{%
}{.}\hspace{.4pt}04\hspace{.1pt}\discretionary{.}{%
}{.}\hspace{.4pt}003}}


\bibitem{drucker2011helping}
S.~M. Drucker, D.~Fisher, and S.~Basu.
\newblock Helping users sort faster with adaptive machine learning
  recommendations.
\newblock In P.~Campos, N.~Graham, J.~Jorge, N.~Nunes, P.~Palanque, and
  M.~Winckler, eds., {\em Human-Computer Interaction -- INTERACT 2011}, pp.
  187--203. Springer Berlin Heidelberg, Berlin, Heidelberg, 2011.

\bibitem{dubey2010clusterlevel}
A.~Dubey, I.~Bhattacharya, and S.~Godbole.
\newblock A cluster-level semi-supervision model for interactive clustering.
\newblock In J.~L. Balc{\'a}zar, F.~Bonchi, A.~Gionis, and M.~Sebag, eds., {\em
  Machine Learning and Knowledge Discovery in Databases}, pp. 409--424.
  Springer Berlin Heidelberg, Berlin, Heidelberg, 2010.

\bibitem{duncan1984selective}
J.~Duncan.
\newblock Selective attention and the organization of visual information.
\newblock {\em Journal of experimental psychology: General}, 113(4):501, 1984.

\bibitem{dunn1974well}
J.~C. Dunn.
\newblock Well-separated clusters and optimal fuzzy partitions.
\newblock {\em Journal of cybernetics}, 4(1):95--104, 1974.

\bibitem{endert2012semantic}
A.~Endert, P.~Fiaux, and C.~North.
\newblock Semantic interaction for sensemaking: Inferring analytical reasoning
  for model steering.
\newblock {\em IEEE Transactions on Visualization and Computer Graphics},
  18(12):2879--2888, Dec 2012. doi: {{%
10\hspace{.1pt}\discretionary{.}{%
}{.}\hspace{.4pt}1109\discretionary{/}{%
}{/}TVCG\hspace{.1pt}\discretionary{.}{%
}{.}\hspace{.4pt}2012\hspace{.1pt}\discretionary{.}{%
}{.}\hspace{.4pt}260}}


\bibitem{endert2012semantics}
A.~Endert, S.~Fox, D.~Maiti, S.~Leman, and C.~North.
\newblock The semantics of clustering: Analysis of user-generated
  spatializations of text documents.
\newblock In {\em Proceedings of the International Working Conference on
  Advanced Visual Interfaces}, AVI '12, pp. 555--562. ACM, New York, NY, USA,
  2012. doi: {{%
10\hspace{.1pt}\discretionary{.}{%
}{.}\hspace{.4pt}1145\discretionary{/}{%
}{/}2254556\hspace{.1pt}\discretionary{.}{%
}{.}\hspace{.4pt}2254660}}


\bibitem{endert2011observation}
A.~{Endert}, C.~{Han}, D.~{Maiti}, L.~{House}, S.~{Leman}, and C.~{North}.
\newblock Observation-level interaction with statistical models for visual
  analytics.
\newblock In {\em 2011 IEEE Conference on Visual Analytics Science and
  Technology (VAST)}, pp. 121--130, 2011.

\bibitem{endert2017state}
A.~Endert, W.~Ribarsky, C.~Turkay, B.~W. Wong, I.~Nabney, I.~D. Blanco, and
  F.~Rossi.
\newblock The state of the art in integrating machine learning into visual
  analytics.
\newblock In {\em Computer Graphics Forum}, vol.~36, pp. 458--486. Wiley Online
  Library, 2017.

\bibitem{estivillcastro2002somany}
V.~Estivill-Castro.
\newblock Why so many clustering algorithms: A position paper.
\newblock {\em SIGKDD Explor. Newsl.}, 4(1):65--75, June 2002. doi: {{%
10\hspace{.1pt}\discretionary{.}{%
}{.}\hspace{.4pt}1145\discretionary{/}{%
}{/}568574\hspace{.1pt}\discretionary{.}{%
}{.}\hspace{.4pt}568575}}


\bibitem{fisher1987improving}
D.~H. Fisher.
\newblock Improving inference through conceptual clustering.
\newblock In {\em AAAI}, vol.~87, pp. 461--465, 1987.

\bibitem{fisher1987knowledge}
D.~H. Fisher.
\newblock Knowledge acquisition via incremental conceptual clustering.
\newblock {\em Machine learning}, 2(2):139--172, 1987.

\bibitem{gillam2001varieties}
B.~Gillam.
\newblock Varieties of grouping and its role in determining surface layout.
\newblock In T.~Shipley and P.~Kellman, eds., {\em From Fragments to Objects:
  Segmentation and Grouping in Vision}, chap.~8, pp. 247--264. Elsevier Science
  B.V., Amsterdam, The Netherlands, 2001.

\bibitem{gower1971general}
J.~C. Gower.
\newblock A general coefficient of similarity and some of its properties.
\newblock {\em Biometrics}, pp. 857--871, 1971.

\bibitem{guo2010interactive}
P.~{Guo}, H.~{Xiao}, Z.~{Wang}, and X.~{Yuan}.
\newblock Interactive local clustering operations for high dimensional data in
  parallel coordinates.
\newblock In {\em 2010 IEEE Pacific Visualization Symposium (PacificVis)}, pp.
  97--104, 2010.

\bibitem{hamilton2014conductor}
P.~Hamilton and D.~J. Wigdor.
\newblock Conductor: Enabling and understanding cross-device interaction.
\newblock In {\em Proceedings of the 32Nd Annual ACM Conference on Human
  Factors in Computing Systems}, CHI '14, pp. 2773--2782. ACM, New York, NY,
  USA, 2014. doi: {{%
10\hspace{.1pt}\discretionary{.}{%
}{.}\hspace{.4pt}1145\discretionary{/}{%
}{/}2556288\hspace{.1pt}\discretionary{.}{%
}{.}\hspace{.4pt}2557170}}


\bibitem{hoque2016interactive}
E.~Hoque and G.~Carenini.
\newblock Interactive topic modeling for exploring asynchronous online
  conversations: Design and evaluation of convisit.
\newblock {\em ACM Trans. Interact. Intell. Syst.}, 6(1), Feb. 2016. doi: {{%
10\hspace{.1pt}\discretionary{.}{%
}{.}\hspace{.4pt}1145\discretionary{/}{%
}{/}2854158}}


\bibitem{ingram2009glimmer}
S.~Ingram, T.~Munzner, and M.~Olano.
\newblock Glimmer: Multilevel mds on the gpu.
\newblock {\em IEEE Transactions on Visualization and Computer Graphics},
  15(2):249--261, March 2009. doi: {{%
10\hspace{.1pt}\discretionary{.}{%
}{.}\hspace{.4pt}1109\discretionary{/}{%
}{/}TVCG\hspace{.1pt}\discretionary{.}{%
}{.}\hspace{.4pt}2008\hspace{.1pt}\discretionary{.}{%
}{.}\hspace{.4pt}85}}


\bibitem{isenberg2008exploratory}
P.~Isenberg, A.~Tang, and S.~Carpendale.
\newblock An exploratory study of visual information analysis.
\newblock In {\em Proceedings of the SIGCHI Conference on Human Factors in
  Computing Systems}, pp. 1217--1226. ACM, 2008.

\bibitem{kang2011how}
Y.~{Kang}, C.~{G{\"o}rg}, and J.~{Stasko}.
\newblock How can visual analytics assist investigative analysis? design
  implications from an evaluation.
\newblock {\em IEEE Transactions on Visualization and Computer Graphics},
  17(5):570--583, 2011.

\bibitem{keil1992concepts}
F.~C. Keil.
\newblock {\em Concepts, kinds, and cognitive development}.
\newblock mit Press, 1992.

\bibitem{kim2016interaxis}
H.~{Kim}, J.~{Choo}, H.~{Park}, and A.~{Endert}.
\newblock Interaxis: Steering scatterplot axes via observation-level
  interaction.
\newblock {\em IEEE Transactions on Visualization and Computer Graphics},
  22(1):131--140, 2016.

\bibitem{klein2006making}
G.~Klein, B.~Moon, and R.~R. Hoffman.
\newblock Making sense of sensemaking 2: A macrocognitive model.
\newblock {\em IEEE Intelligent systems}, 21(5):88--92, 2006.

\bibitem{kopanas2002role}
I.~Kopanas, N.~M. Avouris, and S.~Daskalaki.
\newblock The role of domain knowledge in a large scale data mining project.
\newblock In {\em Hellenic Conference on Artificial Intelligence}, pp.
  288--299. Springer, 2002.

\bibitem{kramer1991perceptual}
A.~F. Kramer and A.~Jacobson.
\newblock Perceptual organization and focused attention: The role of objects
  and proximity in visual processing.
\newblock {\em Perception \& psychophysics}, 50(3):267--284, 1991.

\bibitem{kriegel2009clustering}
H.-P. Kriegel, P.~Kr\"{o}ger, and A.~Zimek.
\newblock Clustering high-dimensional data: A survey on subspace clustering,
  pattern-based clustering, and correlation clustering.
\newblock {\em ACM Trans. Knowl. Discov. Data}, 3(1), Mar. 2009. doi: {{%
10\hspace{.1pt}\discretionary{.}{%
}{.}\hspace{.4pt}1145\discretionary{/}{%
}{/}1497577\hspace{.1pt}\discretionary{.}{%
}{.}\hspace{.4pt}1497578}}


\bibitem{kwon2017axisketcher}
B.~C. {Kwon}, H.~{Kim}, E.~{Wall}, J.~{Choo}, H.~{Park}, and A.~{Endert}.
\newblock Axisketcher: Interactive nonlinear axis mapping of visualizations
  through user drawings.
\newblock {\em IEEE Transactions on Visualization and Computer Graphics},
  23(1):221--230, 2017.

\bibitem{lampert2009animals}
C.~H. Lampert, H.~Nickisch, S.~Harmeling, and J.~Weidmann.
\newblock Animals with attributes: A dataset for attribute based
  classification, 2009.

\bibitem{lee2012ivisclustering}
H.~Lee, J.~Kihm, J.~Choo, J.~Stasko, and H.~Park.
\newblock ivisclustering: An interactive visual document clustering via topic
  modeling.
\newblock {\em Computer Graphics Forum}, 31(3pt3):1155--1164, 2012. doi: {{%
10\hspace{.1pt}\discretionary{.}{%
}{.}\hspace{.4pt}1111\discretionary{/}{%
}{/}j\hspace{.1pt}\discretionary{.}{%
}{.}\hspace{.4pt}1467\discretionary{%
}{-}{-}8659\hspace{.1pt}\discretionary{.}{%
}{.}\hspace{.4pt}2012\hspace{.1pt}\discretionary{.}{%
}{.}\hspace{.4pt}03108\hspace{.1pt}\discretionary{.}{%
}{.}\hspace{.4pt}x}}


\bibitem{lewis2012human}
J.~Lewis, M.~Ackerman, and V.~de~Sa.
\newblock Human cluster evaluation and formal quality measures: A comparative
  study.
\newblock In {\em Proceedings of the Annual Meeting of the Cognitive Science
  Society}, vol.~34, 2012.

\bibitem{lewis2012behavioral}
J.~Lewis, L.~Van~der Maaten, and V.~de~Sa.
\newblock A behavioral investigation of dimensionality reduction.
\newblock In {\em Proceedings of the Annual Meeting of the Cognitive Science
  Society}, vol.~34, 2012.

\bibitem{maaten2008visualizing}
L.~v.~d. Maaten and G.~Hinton.
\newblock Visualizing data using t-sne.
\newblock {\em J. Mach. Learn. Res.}, 9:2579--2605, Sept. 2008.

\bibitem{macinnes2010visual}
J.~{MacInnes}, S.~{Santosa}, and W.~{Wright}.
\newblock Visual classification: Expert knowledge guides machine learning.
\newblock {\em IEEE Computer Graphics and Applications}, 30(1):8--14, 2010.

\bibitem{mander1992pile}
R.~Mander, G.~Salomon, and Y.~Y. Wong.
\newblock A pile metaphor for supporting casual organization of information.
\newblock In {\em Proceedings of the SIGCHI Conference on Human Factors in
  Computing Systems}, CHI '92, pp. 627--634. ACM, New York, NY, USA, 1992. doi:
  {{%
10\hspace{.1pt}\discretionary{.}{%
}{.}\hspace{.4pt}1145\discretionary{/}{%
}{/}142750\hspace{.1pt}\discretionary{.}{%
}{.}\hspace{.4pt}143055}}


\bibitem{mueller2009evaluating}
E.~M\"{u}ller, S.~G\"{u}nnemann, I.~Assent, and T.~Seidl.
\newblock Evaluating clustering in subspace projections of high dimensional
  data.
\newblock {\em Proc. VLDB Endow.}, 2(1):1270–1281, Aug. 2009. doi: {{%
10\hspace{.1pt}\discretionary{.}{%
}{.}\hspace{.4pt}14778\discretionary{/}{%
}{/}1687627\hspace{.1pt}\discretionary{.}{%
}{.}\hspace{.4pt}1687770}}


\bibitem{ng2001spectral}
A.~Y. Ng, M.~I. Jordan, Y.~Weiss, et~al.
\newblock On spectral clustering: Analysis and an algorithm.
\newblock In {\em NIPS}, vol.~14, pp. 849--856, 2001.

\bibitem{nielsen2002getting}
J.~Nielsen, T.~Clemmensen, and C.~Yssing.
\newblock Getting access to what goes on in people's heads?: Reflections on the
  think-aloud technique.
\newblock In {\em Proceedings of the Second Nordic Conference on Human-computer
  Interaction}, NordiCHI '02, pp. 101--110. ACM, New York, NY, USA, 2002. doi:
  {{%
10\hspace{.1pt}\discretionary{.}{%
}{.}\hspace{.4pt}1145\discretionary{/}{%
}{/}572020\hspace{.1pt}\discretionary{.}{%
}{.}\hspace{.4pt}572033}}


\bibitem{nonato2019multidimensional}
L.~G. {Nonato} and M.~{Aupetit}.
\newblock Multidimensional projection for visual analytics: Linking techniques
  with distortions, tasks, and layout enrichment.
\newblock {\em IEEE Transactions on Visualization and Computer Graphics},
  25(8):2650--2673, 2019.

\bibitem{pandey2016towards}
A.~V. Pandey, J.~Krause, C.~Felix, J.~Boy, and E.~Bertini.
\newblock Towards understanding human similarity perception in the analysis of
  large sets of scatter plots.
\newblock In {\em Proceedings of the 2016 CHI Conference on Human Factors in
  Computing Systems}, pp. 3659--3669, 2016.

\bibitem{parsons2004subspace}
L.~Parsons, E.~Haque, and H.~Liu.
\newblock Subspace clustering for high dimensional data: A review.
\newblock {\em SIGKDD Explor. Newsl.}, 6(1):90–--105, June 2004. doi: {{%
10\hspace{.1pt}\discretionary{.}{%
}{.}\hspace{.4pt}1145\discretionary{/}{%
}{/}1007730\hspace{.1pt}\discretionary{.}{%
}{.}\hspace{.4pt}1007731}}


\bibitem{paulovich2011piece}
F.~V. Paulovich, D.~M. Eler, J.~Poco, C.~P. Botha, R.~Minghim, and L.~G.
  Nonato.
\newblock Piece wise laplacian-based projection for interactive data
  exploration and organization.
\newblock In {\em Computer Graphics Forum}, vol.~30, pp. 1091--1100. Wiley
  Online Library, 3 2011.

\bibitem{peck2019data}
E.~M. Peck, S.~E. Ayuso, and O.~El-Etr.
\newblock Data is personal: Attitudes and perceptions of data visualization in
  rural pennsylvania.
\newblock In {\em Proceedings of the 2019 CHI Conference on Human Factors in
  Computing Systems}, CHI ’19. Association for Computing Machinery, New York,
  NY, USA, 2019. doi: {{%
10\hspace{.1pt}\discretionary{.}{%
}{.}\hspace{.4pt}1145\discretionary{/}{%
}{/}3290605\hspace{.1pt}\discretionary{.}{%
}{.}\hspace{.4pt}3300474}}


\bibitem{pirolli1999information}
P.~Pirolli and S.~Card.
\newblock Information foraging.
\newblock {\em Psychological review}, 106(4):643, 1999.

\bibitem{pirolli2005sensemakingprocess}
P.~Pirolli and S.~Card.
\newblock The sensemaking process and leverage points for analyst technology as
  identified through cognitive task analysis.
\newblock {\em Proceedings of International Conference on Intelligence
  Analysis}, 5:2--4, 2005.

\bibitem{robertson1998data}
G.~Robertson, M.~Czerwinski, K.~Larson, D.~C. Robbins, D.~Thiel, and M.~van
  Dantzich.
\newblock Data mountain: Using spatial memory for document management.
\newblock In {\em Proceedings of the 11th Annual ACM Symposium on User
  Interface Software and Technology}, UIST '98, pp. 153--162. ACM, New York,
  NY, USA, 1998. doi: {{%
10\hspace{.1pt}\discretionary{.}{%
}{.}\hspace{.4pt}1145\discretionary{/}{%
}{/}288392\hspace{.1pt}\discretionary{.}{%
}{.}\hspace{.4pt}288596}}


\bibitem{robinson2008collaborative}
A.~C. {Robinson}.
\newblock Collaborative synthesis of visual analytic results.
\newblock In {\em 2008 IEEE Symposium on Visual Analytics Science and
  Technology}, pp. 67--74, Oct 2008. doi: {{%
10\hspace{.1pt}\discretionary{.}{%
}{.}\hspace{.4pt}1109\discretionary{/}{%
}{/}VAST\hspace{.1pt}\discretionary{.}{%
}{.}\hspace{.4pt}2008\hspace{.1pt}\discretionary{.}{%
}{.}\hspace{.4pt}4677358}}


\bibitem{rogers1994distributed}
Y.~Rogers and J.~Ellis.
\newblock Distributed cognition: an alternative framework for analysing and
  explaining collaborative working.
\newblock {\em Journal of Information Technology}, 9(2):119--128, Jun 1994.
  doi: {{%
10\hspace{.1pt}\discretionary{.}{%
}{.}\hspace{.4pt}1057\discretionary{/}{%
}{/}jit\hspace{.1pt}\discretionary{.}{%
}{.}\hspace{.4pt}1994\hspace{.1pt}\discretionary{.}{%
}{.}\hspace{.4pt}12}}


\bibitem{rose1993content}
D.~E. Rose, R.~Mander, T.~Oren, D.~B. Ponc{\'e}leon, G.~Salomon, and Y.~Y.
  Wong.
\newblock Content awareness in a file system interface: Implementing the
  ``pile'' metaphor for organizing information.
\newblock In {\em Proceedings of the 16th Annual International ACM SIGIR
  Conference on Research and Development in Information Retrieval}, SIGIR '93,
  pp. 260--269. ACM, New York, NY, USA, 1993. doi: {{%
10\hspace{.1pt}\discretionary{.}{%
}{.}\hspace{.4pt}1145\discretionary{/}{%
}{/}160688\hspace{.1pt}\discretionary{.}{%
}{.}\hspace{.4pt}160735}}


\bibitem{russell1993cost}
D.~M. Russell, M.~J. Stefik, P.~Pirolli, and S.~K. Card.
\newblock The cost structure of sensemaking.
\newblock In {\em Proceedings of the INTERACT '93 and CHI '93 Conference on
  Human Factors in Computing Systems}, CHI '93, pp. 269--276. ACM, New York,
  NY, USA, 1993. doi: {{%
10\hspace{.1pt}\discretionary{.}{%
}{.}\hspace{.4pt}1145\discretionary{/}{%
}{/}169059\hspace{.1pt}\discretionary{.}{%
}{.}\hspace{.4pt}169209}}


\bibitem{saket2014node}
B.~Saket, P.~Simonetto, S.~Kobourov, and K.~Börner.
\newblock Node, node-link, and node-link-group diagrams: An evaluation.
\newblock {\em IEEE Transactions on Visualization and Computer Graphics},
  20(12):2231--2240, Dec 2014. doi: {{%
10\hspace{.1pt}\discretionary{.}{%
}{.}\hspace{.4pt}1109\discretionary{/}{%
}{/}TVCG\hspace{.1pt}\discretionary{.}{%
}{.}\hspace{.4pt}2014\hspace{.1pt}\discretionary{.}{%
}{.}\hspace{.4pt}2346422}}


\bibitem{sedlmair2012taxonomy}
M.~Sedlmair, A.~Tatu, T.~Munzner, and M.~Tory.
\newblock A taxonomy of visual cluster separation factors.
\newblock {\em Computer Graphics Forum}, 31(3pt4):1335--1344, 2012. doi: {{%
10\hspace{.1pt}\discretionary{.}{%
}{.}\hspace{.4pt}1111\discretionary{/}{%
}{/}j\hspace{.1pt}\discretionary{.}{%
}{.}\hspace{.4pt}1467\discretionary{%
}{-}{-}8659\hspace{.1pt}\discretionary{.}{%
}{.}\hspace{.4pt}2012\hspace{.1pt}\discretionary{.}{%
}{.}\hspace{.4pt}03125\hspace{.1pt}\discretionary{.}{%
}{.}\hspace{.4pt}x}}


\bibitem{self2018observation}
J.~Z. Self, M.~Dowling, J.~Wenskovitch, I.~Crandell, M.~Wang, L.~House,
  S.~Leman, and C.~North.
\newblock Observation-level and parametric interaction for high-dimensional
  data analysis.
\newblock {\em ACM Transactions on Interactive Intelligent Systems (TiiS)},
  8(2):15:1--15:36, June 2018. doi: {{%
10\hspace{.1pt}\discretionary{.}{%
}{.}\hspace{.4pt}1145\discretionary{/}{%
}{/}3158230}}


\bibitem{sellen1997paper}
A.~Sellen and R.~Harper.
\newblock Paper as an analytic resource for the design of new technologies.
\newblock In {\em CHI}, vol.~97, pp. 319--326. Citeseer, 1997.

\bibitem{shipman1999formality}
F.~M. Shipman and C.~C. Marshall.
\newblock Formality considered harmful: Experiences, emerging themes, and
  directions on the use of formal representations in interactive systems.
\newblock {\em Computer Supported Cooperative Work (CSCW)}, 8(4):333--352,
  1999. doi: {{%
10\hspace{.1pt}\discretionary{.}{%
}{.}\hspace{.4pt}1023\discretionary{/}{%
}{/}A\discretionary{:}{%
}{:}1008716330212}}


\bibitem{singhal2001modern}
A.~Singhal et~al.
\newblock Modern information retrieval: A brief overview.
\newblock {\em IEEE Data Eng. Bull.}, 24(4):35--43, 2001.

\bibitem{sourina2005visual}
O.~Sourina and D.~Liu.
\newblock Visual interactive clustering and querying of spatio-temporal data.
\newblock In O.~Gervasi, M.~L. Gavrilova, V.~Kumar, A.~Lagan{\'a}, H.~P. Lee,
  Y.~Mun, D.~Taniar, and C.~J.~K. Tan, eds., {\em Computational Science and Its
  Applications -- ICCSA 2005}, pp. 968--977. Springer Berlin Heidelberg,
  Berlin, Heidelberg, 2005.

\bibitem{tatu2012subspace}
A.~{Tatu}, F.~{Maaß}, I.~{Färber}, E.~{Bertini}, T.~{Schreck}, T.~{Seidl},
  and D.~{Keim}.
\newblock Subspace search and visualization to make sense of alternative
  clusterings in high-dimensional data.
\newblock In {\em 2012 IEEE Conference on Visual Analytics Science and
  Technology (VAST)}, pp. 63--72, 2012.

\bibitem{torgerson1958theory}
W.~S. Torgerson.
\newblock {\em Theory and methods of scaling}.
\newblock Wiley, Oxford, England, 1958.

\bibitem{wenskovitch2018effect}
J.~{Wenskovitch}, L.~{Bradel}, M.~{Dowling}, L.~{House}, and C.~{North}.
\newblock The effect of semantic interaction on foraging in text analysis.
\newblock In {\em 2018 IEEE Conference on Visual Analytics Science and
  Technology (VAST)}, pp. 13--24, 2018.

\bibitem{wenskovitch2018towards}
J.~Wenskovitch, I.~Crandell, N.~Ramakrishnan, L.~House, S.~Leman, and C.~North.
\newblock Towards a systematic combination of dimension reduction and
  clustering in visual analytics.
\newblock {\em IEEE Transactions on Visualization and Computer Graphics},
  24(1):131--141, Jan 2018. doi: {{%
10\hspace{.1pt}\discretionary{.}{%
}{.}\hspace{.4pt}1109\discretionary{/}{%
}{/}TVCG\hspace{.1pt}\discretionary{.}{%
}{.}\hspace{.4pt}2017\hspace{.1pt}\discretionary{.}{%
}{.}\hspace{.4pt}2745258}}


\bibitem{wenskovitch2017observationlevel}
J.~Wenskovitch and C.~North.
\newblock Observation-level interaction with clustering and dimension reduction
  algorithms.
\newblock In {\em Proceedings of the 2nd Workshop on Human-In-the-Loop Data
  Analytics}, HILDA'17, pp. 14:1--14:6. ACM, New York, NY, USA, 2017. doi: {{%
10\hspace{.1pt}\discretionary{.}{%
}{.}\hspace{.4pt}1145\discretionary{/}{%
}{/}3077257\hspace{.1pt}\discretionary{.}{%
}{.}\hspace{.4pt}3077259}}


\bibitem{wenskovitch2019pollux}
J.~Wenskovitch and C.~North.
\newblock Pollux: Interactive cluster-first projections of high-dimensional
  data.
\newblock In {\em 2019 IEEE Visualization in Data Science (VDS)}, pp. 38--47,
  Oct 2019. doi: {{%
10\hspace{.1pt}\discretionary{.}{%
}{.}\hspace{.4pt}1109\discretionary{/}{%
}{/}VDS48975\hspace{.1pt}\discretionary{.}{%
}{.}\hspace{.4pt}2019\hspace{.1pt}\discretionary{.}{%
}{.}\hspace{.4pt}8973381}}


\bibitem{whittaker2001character}
S.~Whittaker and J.~Hirschberg.
\newblock The character, value, and management of personal paper archives.
\newblock {\em ACM Trans. Comput.-Hum. Interact.}, 8(2):150--170, June 2001.
  doi: {{%
10\hspace{.1pt}\discretionary{.}{%
}{.}\hspace{.4pt}1145\discretionary{/}{%
}{/}376929\hspace{.1pt}\discretionary{.}{%
}{.}\hspace{.4pt}376932}}


\bibitem{wilson2002six}
M.~Wilson.
\newblock Six views of embodied cognition.
\newblock {\em Psychonomic bulletin \& review}, 9(4):625--636, 2002.

\bibitem{yi2005dust}
J.~S. Yi, R.~Melton, J.~Stasko, and J.~A. Jacko.
\newblock Dust \& magnet: multivariate information visualization using a magnet
  metaphor.
\newblock {\em Information visualization}, 4(4):239--256, 2005.

\bibitem{young2005direct}
K.~A. Young.
\newblock Direct from the source: The value of 'think aloud' data in
  understanding learning.
\newblock {\em The Journal of Educational Enquiry}, 2005.

\bibitem{zadeh1965fuzzy}
L.~A. Zadeh.
\newblock Fuzzy sets.
\newblock {\em Information and control}, 8(3):338--353, 1965.

\bibitem{zadeh1996note}
L.~A. Zadeh.
\newblock A note on prototype theory and fuzzy sets.
\newblock In {\em Fuzzy Sets, Fuzzy Logic, And Fuzzy Systems: Selected Papers
  by Lotfi A Zadeh}, pp. 587--593. World Scientific, 1996.

\bibitem{zemel2002experience}
R.~S. Zemel, M.~Behrmann, M.~C. Mozer, and D.~Bavelier.
\newblock Experience-dependent perceptual grouping and object-based attention.
\newblock {\em Journal of Experimental Psychology: Human Perception and
  Performance}, 28(1):202, 2002.

\bibitem{zha2002spectral}
H.~Zha, X.~He, C.~Ding, M.~Gu, and H.~D. Simon.
\newblock Spectral relaxation for k-means clustering.
\newblock In {\em Advances in Neural Information Processing Systems}, pp.
  1057--1064, 2002.

\end{thebibliography}
\end{document}